\documentclass[a4paper,conference]{IEEEtran} 
\ifCLASSINFOpdf
  \usepackage[pdftex,dvipdfm]{graphicx}
\else
\fi
\usepackage[cmex10]{amsmath}
\usepackage{amssymb,verbatim}
\usepackage[dvipdfm]{graphicx}
\begin{document}
%
\title{%
Quantum Shannon Information Theory \\
-Design of communication, cipher and sensor-}

\author{
\IEEEauthorblockN{Osamu HIROTA$^{1,2}$ \\}
\IEEEauthorblockA{
1. Quantum ICT Research Institute, Tamagawa University\\
6-1-1, Tamagawa-gakuen, Machida, Tokyo 194-8610, Japan\\
2. Research and Development Initiative, Chuo University, \\
1-13-27, Kasuga, Bunkyou-ku, Tokyo 112-8551, Japan\\
{\footnotesize\tt E-mail:hirota@lab.tamagawa.ac.jp} \vspace*{-2.64ex}}
}

\maketitle

\begin{abstract}

One of the key aspects of Shannon's theory is that it provides guidance for designing the most efficient systems, 
such as minimizing errors and clarifying the limits of coding. Such theories have made great developments 
in the 50 years since 1948. It has played a vital role in enabling the development of modern ultra-fast, stable, 
and highly dependable information and communication systems. The Shannon theory is supported by 
the statistical communication theory such as detection and estimation theory. 
The theory of communication systems that transmit Shannon information using quantum media is called 
quantum Shannon information theory, and research began in the 1960s. 
The theoretical formulation comparable to conventional Shannon theory has been completed. 
Its important role is to suggest that application of quantum effect will surpass existing communication performance. 
It would be meaningless if performance, efficiency, and utility were to deteriorate due to quantum effects, 
even if certain new function is given. This paper suggests that there are various limitations to utilizing 
quantum Shannon information theory to benefit real-world communication systems, and presents a theoretical framework 
for achieving the ultimate goal. Finally, we introduce the perfect secure cipher that overcome 
the Shannon impossibility theorem without degrading communication performance and sensor et al as the examples.

\end{abstract}

%
\IEEEpeerreviewmaketitle

\section{\textbf{Introduction}}
In the 1970s, while upholding the MIT tradition, R.S. Kennedy and H.A. Haus of MIT shifted communications science 
toward a research direction that incorporates fundamental theories of communications that take quantum effects 
into account (Appendix). Kennedy took the lead in this direction and began nurturing young researchers, 
which has resulted in research that has had a global impact in a variety of fields.
The current boom in quantum information science can be traced back to their leadership.

 In this subject, a design theory for systems 
that transmit conventional information using quantum states is called quantum Shannon information theory. 
In order to construct such a theory, probability theory is 
needed to describe quantum mechanical phenomena in an information-theoretic way. 
It is formulated as quantum probability theory, which has a different mathematical system from conventional probability theory and 
can faithfully describe quantum mechanical phenomena. For these reasons, it is necessary to build communication theory 
for the above communication system based on quantum probability theory. It is called 
quantum communication theory (or quantum detection and estimation theory) which were developed by Helstrom [1], Holevo [2],Yuen [3] 
and other researchers.

Like the conventional Shannon theory, the construction of quantum Shannon information theory was started based on 
quantum communication theory, studying of the properties of the Shannon mutual information in systems with quantum effects.
The most important achievement of the evolution from quantum communication theory to quantum Shannon information theory is 
the proof of Shannon's second theorem (channel coding theorem) for quantum communication systems. 
First, Hausladen et al. of the Jozsa group proved it by skillfully using the properties of random coding and quantum measurement 
on a typical subspace of the Hilbert space spanned by pure signal quantum states for the case where the quantum state 
communication channel is noiseless [4]. 

Following from this result, Holevo [5] and Shumacher-Westmoreland [6] proved that the discrete channel capacity of 
a general quantum channel with noise is given by the Holevo information. 
This is now called the Holevo-Shumacher-Westmoreland theorem. 

It is well known that the theorems on the coding limit are the most important in the application of information theory 
among the major theorems in information theory. In particular, the channel coding theorem shows that there exists 
a coding scheme that is error-free for an infinitely long typical sequence, and that the information rate is 
ultimately equal to the channel capacity.

In response to such a limiting existence theorem, Gallager completed a theory that does not use a typical sequence and 
gives a limit to the average error probability of code words for a finite-length code system. This allows us to know 
the limiting characteristics of errors for a finite-length code system for any information rate up to the channel capacity 
when a channel matrix is given. These are called the theory of reliability functions, and it is well known that they made 
a major contribution to the development of the theory in the process of constructing Shannon theory [7].

Thus, as a natural evolution of the research, Holevo's group  introduced a reliability function corresponding to quantum 
communication channels following Gallager's concept [8,9]. In addition, Helstrom, C.Bendjaballa et al [10,11,12] 
and M. Ban et al [13,14] formulated the cutoff rate of quantum systems. These trials are applications of Gallager theory 
to quantum systems. 
Just as Gallager's reliability function theory was effective in analyzing communication systems following classical physics, 
the quantum reliability function and cut off rate are also considered to be effective in analyzing 
communication systems with quantum effects.
Based on this perspective, Holevo has completed a unified mathematical foundation for quantum systems [15].

In this paper, we summarize the development from quantum communication theory to quantum Shannon information theory through 
the section II $\sim$ the section V and explain its applications to real-world communication technologies, 
showing quantum advantage for communication, cipher and sensors.
The most dramatic result of this theory is the discovery of quantum stream ciphers, which lift the Shannon impossibility theorem 
for symmetric key cipher and realize perfectly secure ciphers with short keys.

\section{\textbf{General quantum communication channel model}}
In this section, we denote the theoretical structure of general quantum systems and provide an explanation of 
the mathematical theory of quantum communication channel. In quantum theory for transmission of the information 
defined by Shannon, Shannon's information as a finite set of alphabet is mapped to a quantum state. 
Therefore, the overall communication channel consists of a quantum state to quantum state communication channel 
and a quantum measurement channel that converts the received quantum state into a classical signal via measurement (Fig.1).

In order to analyze them as information theory, it is necessary to set up a communication channel model. 
A communication channel that transmits a quantum state onto which signal is mapped is called a quantum state 
transmission channel. This channel is described by a completely positive map in the general mathematical sense [15].\\

$\textbf {Definition 1}:$
Let ${\cal A}, {\cal B}$ be two *-algebras and ${\bf S} \subset {\cal A}$ an operators system. 
A linear map $\Phi : {\bf S} \rightarrow {\cal B}$ is called completely positive if
$\sum^{n}_{i,j=1} b^*_i \Phi(a_i^* a_j)b_j \in {\cal B}$ is positive for all $n \in {\bf N}$ and for all 
$a_i \in {\bf S}, b_i \in {\cal B}$.
The space of all such maps is denoted by ${\rm CP}({\bf S}, {\cal B})$.\\

The key issue is the physical reality of such abstract descriptions.
In Shannon theory, a channel is considered noiseless when the relationship between the input and output symbols is injective. 
From this perspective, it is important that the quantum state carrying Shannon information does not lose its quantum properties 
along the channel. Fortunately, the following theorem has been shown for real-world channels by Helstrom [16].\\

\textbf {Theorem 1}:$\{ Helstrom \}$\\
In the transmission of quantum states through an optical communication channel with energy loss, the quantum state 
that can maintain a pure state is a coherent state which is definded by $\textbf{a}|\alpha > = \alpha |\alpha>$, 
where ${\bf a}$ is photon annihilation operator.\\

This theorem has an affinity with modern communication technology.
That is, the maximum capacity of the reliable information transmission in the general optical channeles such as fiber and free space 
is achieved by coherent state, under considering the channel characteristics, bandwidth of light source and receiver.

\begin{figure}
\centering{\includegraphics[width=8cm]{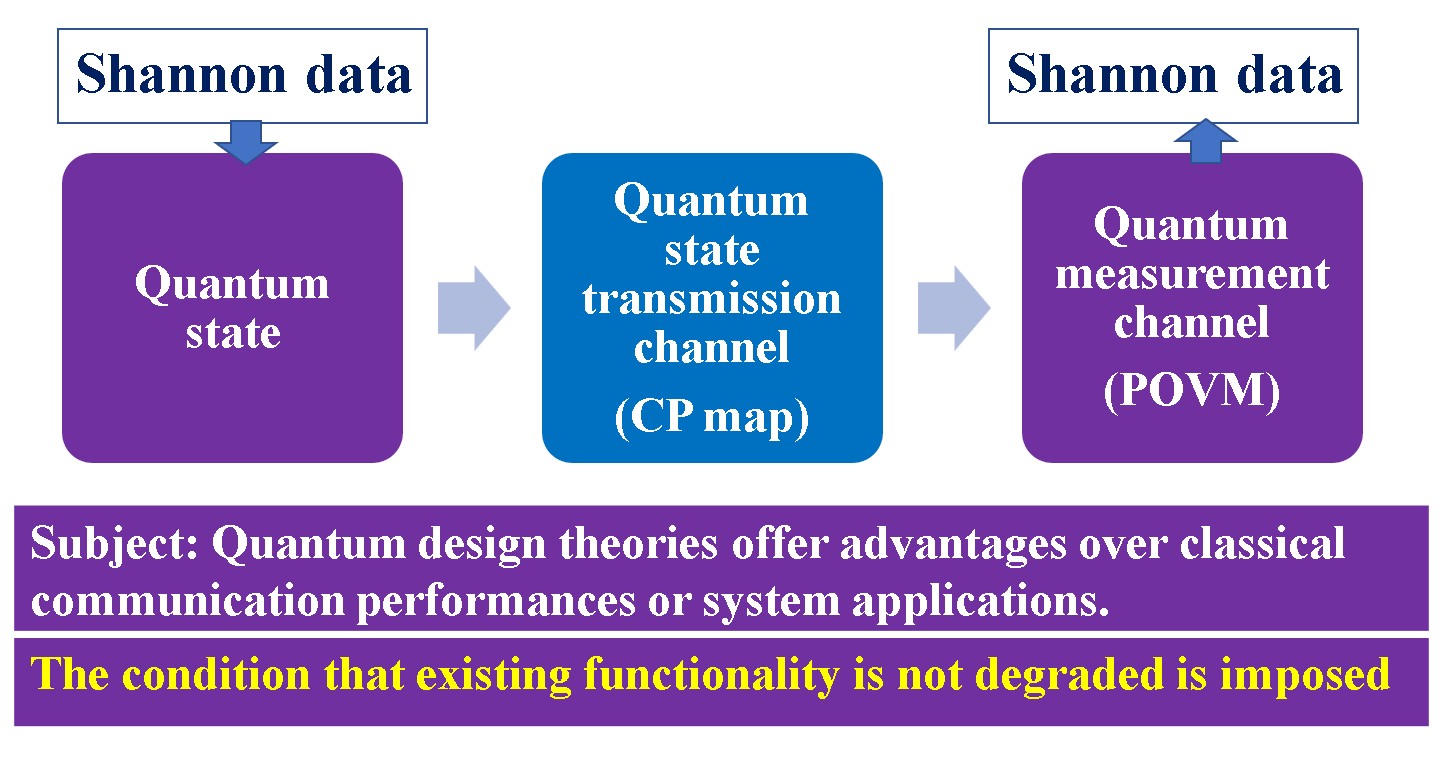}}
\caption{Channel model in quantum Shannon information theory. Shannon information such as digital data or analog signal is 
transmitted by optical signal with quantum effect governed by quantum state. 
The quantum state and quantum measurement determine the communication performance. The existing functionality should not be degraded.}
\end{figure}

\begin{figure}
\centering{\includegraphics[width=8cm]{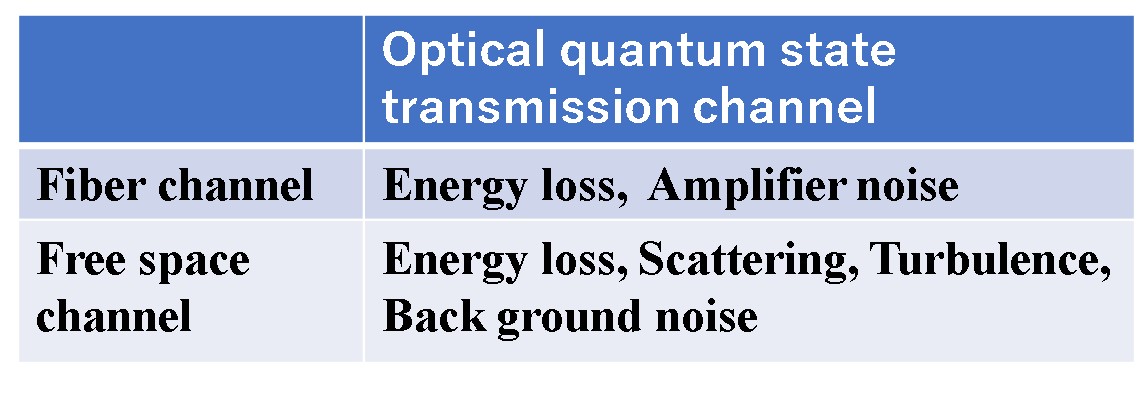}}
\caption{Subject of consideration of quantum state transmission channel. 
The performance requirements for fiber optic communication are 100Gbit/sec to 100Tbit/sec 
as the transmission speed and a communication distance of 10,000km. 
For terrestrial spatial transmission, the speed is 1Gbit/sec to 10Gbit/sec and the communication distance is about 10km.
There are no quantum states other than coherent states that satisfy these requirements. This is a consequence of Theorem 1.}
\end{figure}

In fact, in optical fiber communication, energy loss is the main issue. When the transmitting quantum state is a coherent state, 
which is the quantum state of general laser light, it maintains a pure state even if there is any loss, 
so it is treated as noiseless in the sense of quantum state transmission.

In free-space optical communication, in addition to energy loss, external noise is added, so the quantum state transmission channel 
becomes a  more noisy channel (Fig.2). But still the coherent state provides the maximum capacity which is discussed 
in the section V.\\

$\textbf {Remark 1} :$\\
When single photon (number state), squeezed state, and quantum entangled states, etc. are used as the transmiter state, 
energy loss or decoherence transforms these states to a mixed state and the channel becomes non injective. 
Thus, light sources of quantum states with strong quantum properties cannot meet the performance requirements 
of modern communications which requires over 100 Gbit/sec.\\

Next, it is necessary to describe the process of identifying the quantum state that has arrived at the receiver. 
In this quantum measurement process, unavoidable noise due to the most basic observation operation in quantum mechanics appears. 
If this is modeled as quantum noise, it can be modeled as a quantum measurement communication channel. 
In other words, unavoidable errors occur in the identification of quantum states, and an optimization problem arises 
regarding their decision. 

This type of model was proposed by Helstrom in 1967 and is called the quantum detection and estimation theory [1]. 
The noise model of the photoelectric conversion process in the receiver of laser optical communication is 
the simplest physical example of quantum noise. In order to describe this quantum measurement channel, 
quantum probability theory, which describes the quantum measurement process, is required. 
Below we briefly present the mathematics of quantum probability as a basis for the following discussion.\\

\section{\textbf{Mathematical foundation of generalized quantum measurement and decision operator}}

\subsection{\textbf{Mathematical Formulae of Generalized Quantum Measurement}}

The mathematical structure of quantum mechanics is given based on the Hilbert space theory constructed by von Neumann. 
The representation of the quantum measurement process is formulated using eigen state and eigen value of a self adjoint operator.
That is, it is defined as follows.\\

$\textbf {Definition 2}:$
Let us assume that the quantum system has the self adjoint operator as physical observable  
${\bf T}$ and its quantum state $\rho $. 
The Born rule in the standard quantum measurement of the observable is described as follow:
\begin{equation}
{\bf T}|x >=x|x >, P(x) = Tr \rho | x >< x |
\end{equation}
where $| x >< x |$ is called projection valued measure.\\

The set of density operator $\rho$ is a convex set, its extreme points being the one-dimensional projection. 
The corresponding states are called pure. Any measurement with values in the real number space $R_M$ is described 
by an affine map of the set of the states (density operator) into the set of probability distributions on $R_M$.

However, P. Benioff and others have been discussing the mathematical generalization of the projection process to describe 
the diversity of measurement processes. Finally the generalized quantum measurement process is described by the following form [17].\\

 $\textbf {Definition 3}:$
Let us consider a generalized resolution of identity $\{{\bf X}(r) ; r \in R_M \}$, i.e., the collections of Hermitian operators, 
satisfying $\sum_{r\in R_M}{\bf X}(r) =I, {\bf X}(r) \ge 0$.
$Tr \rho {\bf X}(r)$ establishes the one-to-one correspondence between affine maps of the set of density operators into
 the set of probability distributions on $R_M$ and the resolution of identities.
In some cases, ${\bf X}(r)$ is called the positive operator valued measure (POVM). \\
 
 This formulation makes it possible to formulate the Born effect in quantum measurements 
 without going through concrete physical quantities (observable).

\subsection{\textbf{Quantum Decision Operator}}
In the quantum communication theory, the POVM is used as decision operator : $\{\Pi_m\}$ by following operational correspondence.

Let us assume the set $\rho_m, m=\{1,2,3,\dots, M\}$ of quantum states.
 The decision result through the quantum measurement is described by a compact set of the decision operator: 
$\Pi_l$, $l=1,2,3, \dots ,M$ such that 

\begin{eqnarray}
P(l|m)&=&Tr \rho_m \Pi_l, \quad m,l=1,2,3,\dots, M \\
\sum^{M}_{l=1} \Pi_l &=&I, \quad \Pi_l \ge 0 \quad \forall l
\end{eqnarray}
where $I$ is the identity operator.\\

$\textbf {Remark 2} :$
The decision operator does not just represent the probability of a quantum measurement process, but the probability of 
error or detection by the receiver's decision.
 In other words, it must be understood that it involves a decision by observer [1].\\
 
 On the above remark, from the origins of Helstrom's formulation, it is easier to understand if we interpret 
 the quantum decision operator as the generalization of  the Wald's decision function in the classical system [18].

In the classical communication theory, the decision is applied to the given probability function to variable of received signal. 
When one applies the standard quantum measurement, the probability function is given and the decision function is 
applied to its probability function same as the classical detection theory.
However, in quantum case, the decision operator does not need the explicit probability function of the variable 
of the received signal.
The decision operator directly outputs the probability value of the correctness or incorrectness of the decision 
without going through the probability function of the measurement process.
As an effect of the above fact, the possibility arises that the discrimination capability  to quantum signal
in the quantum measurement process will exceed the discrimination capability by the standard liklihood decision 
based on standard quantum measurement and its probability function. 
{\it \textbf{This is equivalent to violating communication-theoretical causality}}[19, 20].

However, it can enjoy only in the discrete signal set.
 Its effect disappears in the case of continuous variable. That is, in the quantum estimation theory, 
there is no such quantum advantage. These will be discussed in the subsequent sections.

\subsection{\textbf{Decision Operator based on Entangled Measurement (or collective measurement)}}

As mentioned above, a quantum measurement process is considered as a communication channel. In information theory, 
information is generally transmitted as a codeword of length $n$. When extended to $n$-th degree, 
quantum communication channels need to be treated differently from conventional extension channels.

Let us consider a finite alphabet $\{a_l\}, l=(1,2,\dots, M)$ with quantum state $\rho_l$. 
The code word ${\bf k}$ is a combination of the alphabet $\{a_l\}$ with the length $n$.
That is, the input code ${\bf k}=(k_1, k_2,\dots,k_i, \dots,  k_n)$ and output is ${\bf j}=(j_1, j_2,\dots,j_i, \dots, j_n)$, 
where $i$ is the order of sequence in the code, and $0 \le k_i \le M$ and $0 \le j_i \le M$. 
The compaund state $\rho_{\bf k}=(\rho_{k_1}\otimes \rho_{k_2} \dots \otimes \rho_{k_n})$ are 
in a tensor product Hilbert space $H^{\otimes n}=H \otimes \dots \otimes H$.
When the $n$-th extended signal state is used, the operation of treating a sequence of quantum states of length $n$ 
as a single quantum state in the $n$-th extended Hilbert space is called collective or entangled measurement based decoding.
 In this case, these structures provide unique performance in the quantum Shannon information theory, 
 which will be described in the next section. 
In cases of individual and entangled measurement, the optimizing the decision operator that represents 
the decision process is an important issue.\\

\section{\textbf{Structure of quantum detection and estimation theory}}
\subsection{\textbf{Quantum Detection Theory}}
\subsubsection{\textbf{Basic formula}}

The appearance of quantum effects in measurement process of signals and the result of the decision are 
simultaneously characterized 
by the quantum descision operator. 
Thus, the quantum Bayes rule is formulated without going through the likelihood ratio as follows:
\begin{equation}
{\bar {P}_e}=\min_{\{\Pi \}}\{1-\sum_{m=1}^{M} \xi_m Tr \rho_m \Pi_m\}
\end{equation}
where a priori probability must be $(\xi_m >0, \quad \forall m)$ for the admissibility in the decision theory.
The necessary and sufficient condition for $\{\Pi_m\}$ are given by Holevo [2] and Yuen [3]:\\

$\textbf {Theorem 2}$ $\{Holevo, Yuen\}$:
The necessary and sufficient condition for $\{\Pi_m\}$ on the quantum Bayes rule is given by 
\begin{eqnarray}
&&\Pi_m[\xi_m \rho_m -\xi_l \rho_l]\Pi_l =0, \quad \forall l,m \nonumber \\
&&\gamma -\xi_l \rho_l \ge 0, \quad \forall l \nonumber \\
&&\gamma =\sum_l \xi_l \rho_l \Pi_l
\end{eqnarray}

On the other hand, following the classical detection theory, the quantum minimax rule for a signal system 
in which the functional with respect to $\{\xi_m\}$ is a convex function is defined as follows:

\begin{equation}
{\bar {P}_e}=\max_{\{\xi\}} \min_{\{\Pi\}}\{1-\sum_{m=1}^{M} \xi_m Tr \rho_m \Pi_m\}
\end{equation}
and the necessary and sufficient conditions are as follows: [21]:\\

$\textbf {Theorem 3}$ $\{Hirota \cdot Ikehara\}$:
The necessary and sufficient condition for $\{\Pi_m\}$ on the quantum minimax rule is given by
\begin{eqnarray}
&&Tr \Pi_l \rho_l = Tr \Pi_m \rho_m, \quad \forall l,m \nonumber \\
&&\Pi_m[\xi_m \rho_m -\xi_l \rho_l]\Pi_l =0, \quad \forall l,m \nonumber \\
&&\gamma -\xi_l \rho_l \ge 0, \quad \forall l \nonumber \\
&&\gamma =\sum_l \xi_l \rho_l \Pi_l  
\end{eqnarray}

For further formulae please see the references.\\

\subsubsection{\textbf{Useful analytical issues}}

In general, it is very difficult to find the solutions of the above two quantum detection rules.
However, when the signal system satisfies certain conditions or the decision operator can be set in advance, 
the optimal theory becomes very simple.
Some examples of these are described below.

A set of quantum states that conforms to the following definition is particularly convenient.\\

$\textbf {Definition 4}$:\\ 
Let $G$ be a group with an operation $\circ$.
The set of quantum state signals is called group covarinat if there exist unitary operators $U_k(k\in G)$ such that
\begin{equation}
U_k|\psi_m >=|\psi_{k\circ m}>, \forall m,k \in G
\end{equation}
It characterizes quantum states $\{|\psi_m>, m\in G\}$ [17].\\

Let us explain the case of setting a decision operator in advance. Belavkin [22] introduced the following decision operator 
called square root measurement (SRM).\\

$\textbf {Definision 5}$\\
The decision operator called a square root measurement (SRM) is given by .
\begin{eqnarray}
&&\Pi_l =|\mu_l><\mu_l|  \\
&&|\mu_l>= {\Gamma}^{-1/2}|\alpha_l> \nonumber \\
&&{\Gamma}= \sum_{m=1}^M |\alpha_m><\alpha_m| \nonumber
\end{eqnarray}
where $\Gamma$ is a Gram operator.

Based on the above formula, the general properties of quantum Bayes rule  and quantum minimax rule based on 
the above decison operator are given by Ban and Osaki [23,24].\\

\subsubsection{\textbf{Decision operator based on SRM of entangled measurement}}
In the case of the operation of individual decision to the quantum states corresponding to each slot of a codeword, 
the decision operator and decision probability are given by
\begin{eqnarray}
\Pi_{\bf j}&=&\Pi_{j_1}\otimes \Pi_{j_2} \dots \otimes \Pi_{j_n} \\
P({\bf j}|{\bf k})&=& P(j_1|k_1)\times  P(j_2|k_2)\times \dots \times  P(j_n|k_n)\nonumber
\end{eqnarray}
where $P(j_l|k_l)=Tr\rho_{j_l}\Pi_{j_l}$, for $l=\{1,2,\dots, n \}$.
In this case, there is no correlation among the signal decision process. 

On the other hand, one can define the decision operator based on the entangled measurement.
Let us denote the quantum state of $n$-th extended quantum state as follows:
\begin{equation}
|\Psi_{\bf k} >=|\psi_{k_1}>|\psi_{k_2}\dots |\psi_{k_l}>\dots |\psi_{k_n}>
\end{equation}
where $k_l=\{1,2,\dots, M\}, \forall l$.
Then we can adopt the square root measurement as follows:
\begin{eqnarray}
&&\Pi_{\bf j} = |\Phi_{\bf j}><\Phi_{\bf j}|,\\
&&|\Phi_{\bf j}>=(\sum_{\bf k} |\Psi_{\bf k} ><\Psi_{\bf k}|)^{-1/2} |\Psi_{\bf j} >\nonumber\\
&&P({\bf j}|{\bf k})=Tr \rho_{\bf k}\Pi_{\bf j}\nonumber
\end{eqnarray}
where $\rho_{\bf k}=|\Psi_{\bf k}><\Psi_{\bf k}|$. This is called decision operator based on entangled measurement. 
In this case, in addition to the quantum effects due to the decision action on each alphabet, 
quantum effects due to entanglement can be expected.\\

\subsubsection{\textbf{Quantum advantage}}
The performance of optical communications, which is designed using classical communication theory, is determined 
by the performance of existing receivers such as photon counting, homodyne and heterodyne. 
If the performance of a communication system designed using quantum theory exceeds 
conventional performance, it is called quantum advantage or quantum gain. 
The quantum gain characteristics by the quantum decision operator are exemplified below.
The concrete advantage of decision operator based on entangled measurement is discussed in the section V.
\\

$\textbf {Theorem 4}$\\
If the signal set $\{|\alpha_m>\}$ is a covariant, the optimum decision operator is given by the decision operator 
based on the square root measurement.  
And the error probability of quantum Bayes and quantum minimax rules for $M$ covariant signals can be given as follows :
\begin{eqnarray}
{\bar P}_e &=&1- \frac{1}{(M)^2}(\sum_{m=1}^{M} {\sqrt \lambda_m})^2  \\
\lambda_m&=&\sum_{k=1}^{M} <\alpha_1|\alpha_k >u^{-(k-1)m} \nonumber
\end{eqnarray}
where $u=\exp[\pi i/M]$.\\

Here we show the simplest example. Assuming the signal system is binary PSK $\rho_1=|\alpha><\alpha|, 
\rho_2=|-\alpha><-\alpha|$ with $\xi_1=\xi_2=1/2$, 
the quantum solution is as follows [1]:
\begin{eqnarray}
{\bar P}^{opt}_e &=& 1-(Tr\rho_1\Pi^{opt}_1 +Tr\rho_2\Pi^{opt}_2) \nonumber \\
&=&\frac{1}{2}[1- \sqrt{1-|<\alpha|-\alpha>|^2}] \nonumber \\
&\ll& {\bar P}_e(Homodyne) 
\end{eqnarray}
This is called Helstrom bound. The decision operator in this case is given as follows:
\begin{equation}
\Pi^{opt}_1=|\omega_1 ><\omega_1|, \quad \Pi^{opt}_2=|\omega_2 ><\omega_2|
\end{equation}
where
\begin{eqnarray}
|\omega_1>&=&\sqrt{\frac{1+\sqrt{1-\kappa^2}}{2(1-\kappa^2)}}|\alpha >\nonumber \\
&-& \sqrt{\frac{1-\sqrt{1-\kappa^2}}{2(1-\kappa^2)}}|-\alpha > \nonumber \\
|\omega_2>&=&\sqrt{\frac{1-\sqrt{1-\kappa^2}}{2(1-\kappa^2)}}|\alpha > \nonumber \\
&-& \sqrt{\frac{1+\sqrt{1-\kappa^2}}{2(1-\kappa^2)}}|-\alpha >
\end{eqnarray}
where $\kappa=|<\alpha|-\alpha>|$.

In the case of homodyne receiver (classical solution), the measurment operator is 
\begin{equation}
\Pi(x)=|x >< x |, \quad X_c|x>=x|x>, \quad X_c=\frac{1}{2}({\bf a}+{\bf a}^{\dagger})
\end{equation}
The average error probability is given by likelihood test based on $P(x|\alpha)=Tr\rho_1 \Pi(x), 
P(x|-\alpha)=Tr \rho_2 \Pi(x)$. 
It corresponds to the post measurment procedure.
The result does not show the quantum advantage.

Separate analysis is required to determine what physical process the quantum optimal solution Eq(16) corresponds to.
An example is the Dolinar receiver [1].

\subsection{\textbf{Quantum Estimation Theory}}

The quantum estimation theory was also pioneered by Helstrom, Holevo, Personick  and Yuen. 
Then M.G.A. Paris et all discussed many aspect of its application [25].
 In this section, we show the formulation and discuss certain problem.\\

\subsubsection{\textbf{Formulation}}
When physical signal parameter is continuous, a set of the quantum states can be treated as a quantum state system 
corresponding to a continuous signal. The optimization theory goes to quantum estimation from quantum detection theory.
Therefore, the decision must deal with a continuous quantity $\{x \}$. So for the density operator $\rho(x)$, 
the decision is defined as infinitesimal operator.
\begin{eqnarray}
&&\{\Pi_i\} \rightarrow \{ d\Pi(x) \} \\
&&P(x_{out}|x_{in})=Tr \rho(x_{in}) d\Pi(x_{out})\nonumber 
\end{eqnarray}
where $x=x_{in}=x_{out}$ for the correct probability.
Thus, quantum Bayes estimation can be formulated by replacing the discrete system with an infinitesimal decision operator. 
In this case, the optimum POVM corresponds to the measurement process of the observable as the parameter. 
That is, the optimum operator is given by the eigenstate of the operator of physical observable.
Thus, the estimation mechanism will use the maximum likelihood method, just like in classical methods, 
based on $P(x)=Tr\rho(x) d\Pi(x)$.
Hence, it is easy to understand that it has no quantum advantage in the decision process. 

Alternatively, one can adopt the minimum mean square error(MMSE) as the evaluation function.
Personick formulated MMSE estimator which is applicable to linear filtering of random signal sequence [26]. 
On the other hand, Cramer-Rao bound is formulated by Helstrom as follows [1]:\\

$\textbf {Theorem 5}$ : 
The estimation bound for single parameter is given by following formula.
\begin{equation}
Var {\hat x} =Tr \rho ({\hat x} - {x})^2 \ge \frac{1}{Tr \rho(x) {L}_S^2} 
\end{equation}
where $L_S$ is called symmetric logarithm derivative, and is given by
\begin{equation}
\frac{\partial \rho(x)}{\partial x}=\frac{1}{2}[\rho(x){L}_S +{ L}_S \rho(x)]
\end{equation}
The equality holds in the case as follows:
\begin{equation}
{L}_S=k(x)(\hat {{\bf x}}-{ x})
\end{equation}
where ${\hat x}$ is estimate, $k(x)$ is a function of $x$, and the estimate operator ${\hat {\bf x}}$ is 
given by the operator corresponding to the parameter.\\

It means that the optimum measurement is the standard quantum measurement,
and that there is no quantum advantage at the measurement process itself. 
In addition, it has still certain difficulty as denoted in the following.\\

\subsubsection{\textbf{Example for coherent state signal}}
In optical communications and optical radar, signals generally have complex amplitudes (quadrature amplitude):
\begin{equation}
{\alpha}^s =x^s_c + ix^s_s =|{\alpha}^s|\cos \theta + i |{\alpha}^s|\sin \theta =\sqrt {N_s}e^{i\theta}
\end{equation}
Its quantum counterpart is photon annihilation operator ${\bf a} =X_c +i X_s$, where
\begin{equation}
X_c =\frac{1}{2}({\bf a} + {\bf a}^{\dagger}), \quad X_s=\frac{1}{2i}({\bf a} - {\bf a}^{\dagger})
\end{equation}
Here $X_c$ and $X_s$ are non commutative observable. 
The problem is how to formulate the simultaneous estimation for such non commutative observable.\\

(a) \textbf{Single parameter estimation}\\
In general, the density operator of optical field is given as follows:
\begin{equation}
\rho=\frac{1}{\pi N}\int \exp\{-\frac{|\alpha - \alpha^s|^2}{N}\} |\alpha><\alpha|d^2\alpha 
\end{equation}
where $N$ is a back ground noise energy and $\alpha^s$ is the signal. 
When we consider the estimation of the single parameter: $x^s_c$ of $\alpha^s=x^s_c + ix^s_s$. 
Then the solution is given as follows:
\begin{eqnarray}
\frac{\partial  {\rho}}{\partial  x^s_c}&=&\frac{1}{2}({\rho}{L}_S +{ L}_S {\rho})\\
{L}_S&=&\frac{2}{N+1/2}(\frac{{\bf a} + {\bf a}^{\dagger}}{2} -x^s_c)
\end{eqnarray}
As a result, we have 
\begin{equation}
{\hat {\bf x}}= X_c =\frac{1}{2}({\bf a} + {\bf a}^{\dagger}), \quad 
 Var \hat {x}_c = \frac{1}{2}N + \frac{1}{4}
\end{equation}
where $1/4$ means the effect of quantum noise. 
The estimate operator ${\hat {\bf x}}$ means a homodyne receiver. Thus, it has no quantum advantage.\\

(b) \textbf{Non commutative parameter estimation}\\
Let us consider the simultaneous estimation of the non commutative observable :$X_c$ and $X_s$. This is important for the evaluation of 
optical phase estimation such as 
\begin{equation}
\theta =\tan^{-1} \frac{x_s}{x_c}
\end{equation}

In this case, the theorem 5 does not provide the correct solution for the design of communication system.
The quantum optimum measurement for the noncommutative quadrature amplitude can be given
by Yuen-Lax quantum Cramer-Rao bound [27].\\

$\textbf {Theorem 6}$ : 
The estimation bound for complex amplitudes are given by following formula.
\begin{equation}
Var ({\hat \alpha}) \ge \frac{1}{Tr \rho L_R L_R^{\dagger} }
\end{equation}
where the right logarithm derivative $L_R$ is defined by 
\begin{equation}
\frac{\partial \rho}{\partial \alpha^s} = L_R^{\dagger} \rho, \quad L_R = k({\alpha}^s)({\bf a}-{\alpha}^s)
\end{equation}
where ${\bf a}$ is a photon annihilation operator, and it corresponds to a heterodyne measurement. 
Then the bound is given as follows:
\begin{equation}
 Var \hat {\alpha}=N + 1
\end{equation}
where $1=1/2 +1/2$ is the similtaneous measurement effect of the non commutable observable.\\

If one adopts the separate measurement of the quadrature amplitude by two balanced homodyne receivers.
From Theorem 1, the coherent state maintains the coherent state at the output of the half mirror.
Here each measurement may adopt the theorem 5.
\begin{eqnarray}
\frac{\partial \rho(x^s_c)}{\partial x^s_c}&=&\frac{1}{2}[\rho(x^s_c){ L}_S +{L}_S\rho(x^s_c)] \\
\frac{\partial \rho(x^s_s)}{\partial x^s_s}&=&\frac{1}{2}[\rho(x^s_s){L}_S +{L}_S \rho(x^s_s)]
\end{eqnarray}
Each bound is given by 
\begin{equation}
 Var \hat {x}_c = \frac{1}{2}N + \frac{1}{4}, \quad  Var \hat {x}_s = \frac{1}{2}N + \frac{1}{4}
\end{equation}
However, one should not ignor the fact that the signal energy is half at each branch. Thus, the signal to noise ratio for quantum part is 
\begin{eqnarray}
SNR(x_c) &=& \frac{N_s/2}{1/4}=\frac{N_s}{1/2}, \nonumber \\
SNR(x_s) &=& \frac{N_s/2}{1/4}=\frac{N_s}{1/2},
\end{eqnarray}
This means that a set of two balanced homodyne receiver systems is equivalent to heterodyne receiver for the coherent state signal.
The theorem 6 includes the above effect under fixed energy conditions.

Now consider the case when the signals are other than coherent state. In separate measurement of non commutative parameters,  
the quantum state is disturbed by a loss of half mirror.
Therefore, it is not accurate to evaluate communication performance based on the conventional quantum Cramer-Rao bound only.\\

\subsubsection{\textbf{Application of Lie algebra for non commutative parameters}}

We have been studying the theory of quantum state control and simultaneous measurement of non-commutative quantities based on Lie algebra.
Here we introduce it and its application to the formulation of estimation theory. First, we define the formulation as follows:\\

$\textbf {Definition 6}$: Algorithm: \\
Step 1: Derive the estimation operators for simultaneous estimation of non-commutative quantities based on 
the symmetric logarithmic differential operator formula Eq(32),and Eq(33).\\
Step 2: Construct a decision operator that expresses simultaneous measurement of noncommuting quantities 
based on the minimum uncertain state for the two operators obtained.\\

(a) \textbf{Construction of decision operator}. \\

$\textbf {Definition 7}$ : 
An algebra is called a Lie algebra when its multiplication satisfies the following two conditions:
\begin{eqnarray}
&&[{\bf A}, {\bf A}]=0, \\
&&[[{\bf A}, {\bf B}], {\bf C}] +[[{\bf B}, {\bf C}], {\bf A}] + [[{\bf C}, {\bf A}], {\bf B}]=0 \nonumber
\end{eqnarray}
where $[\quad ]$ is the commutation relation.\\

An example of its application in quantum mechanics is the Heizenberg-Weyl group: $W_1$. 
The unitary representation of the sub group of $W_1$ is 
\begin{equation}
{\bf T}({\bf g})=\exp\{is\}\exp\{\alpha{\bf a}^{\dagger} -\alpha^*{\bf a}\}
\end{equation}
where ${\bf g}$ is the elements of Lie algebra.\\

$\textbf {Definition 8}$ : 
For any element $|\psi >$ of Hilbert space, the quantum state constructed by the following formula 
\begin{equation}
{\bf T}({\bf g})|\psi >=|\alpha_{\psi} >_G
\end{equation}
is called generalized coherent state.\\

Here let us consider non-compact groups in Lie algebras.
With respect to Hermite forms, a set that constitute linear isometric groups is non-compact group.
SU(1,1) is a typical example.

The Lie algebra of SU(1,1) has the following elements :${\bf K}_1, {\bf K}_2, {\bf K}_0$. Their commutation relations are
\begin{eqnarray}
&&[{\bf K}_1, {\bf K}_2]=-i{\bf K}_0, [{\bf K}_2, {\bf K}_0]=i{\bf K}_1 \nonumber \\
&&[{\bf K}_0, {\bf K}_1]=i{\bf K}_2
\end{eqnarray}
Then we define ${\bf K}_{\pm}={\bf K}_1 \pm i{\bf K}_2$.
The unitary representaion of SU(1,1) is given by 
\begin{equation}
{\bf T}^{\tau}({\bf g})=\exp\{\zeta {\bf K}_+ - \zeta^* {\bf K}_-\}
\end{equation}
where $\tau$ is given by the property of Casimir operator. The the following state is also the generalized coherent state.
\begin{equation}
{\bf T}^{\tau}({\bf g})|\psi > =|\alpha_{\tau} >_G
\end{equation}
where $|\alpha_{\tau} >$ is called a base state.
When the above state satisfies the minimum uncertainty relation for two observales, 
it is called  the ideal generalized coherent state.

Here for the quadrature amplitude, the elements of SU(1,1) are as follows:.
\begin{eqnarray}
&&{\bf K}_+=\frac{1}{2}{{\bf a}^{\dagger}}^2,  {\bf K}_-=\frac{1}{2} {\bf a}^2, \nonumber \\
&& {\bf K}_0=\frac{1}{4}({\bf a}{\bf a}^{\dagger} + {\bf a}^{\dagger}{\bf a})
\end{eqnarray}
Then we have 
\begin{eqnarray}
&& {\bf T}^{\tau}({\bf g})=\exp\{\frac{z}{2}({\bf a}^2 - {{\bf a}^{\dagger}}^2)\} \\
&&{\bf T}^{\tau}({\bf g}){\bf a} {{\bf T}^{\tau}}^{\dagger}({\bf g})=\mu {\bf a} + \nu {\bf a}^{\dagger} \equiv {\bf b}\\
&& |\alpha_{\tau} > = | 0 >, \quad |\mu|^2 - |\nu|^2=1  \nonumber
\end{eqnarray}
Then the following state is the ideal generalized coherent state.
\begin{eqnarray}
&&{\bf b} |\beta, \mu, \nu > = \beta |\beta, \mu, \nu > \\
&&{\bf T}^{\tau}({\bf g}) |0 > = |0, \mu, \nu >
\end{eqnarray}
where $\beta = \mu \alpha + \nu \alpha^*= {\bar x}^G_c + i {\bar x}^G_s$. 
Eq(45) and Eq(46) give the minimum uncertainty state for $x_c$ and $x_s$.

Let us define the decision operator as the step 1:
\begin{equation}
d\Pi(x_c, x_s) = |\beta, \mu, \nu ><\beta, \mu, \nu |
\end{equation}
This corresponds to the generalized heterodyne receiver [28]. 
Let us show how to use the above formula in the following. \\

(b) \textbf{Estimation bound}. \\
Using these formulas, the modified quantum Cramer-Rao bound for the simultaneous estimation of non commutative parameters 
of coherent state signal without background noise is given by the step 2  as follows [28]
\begin{eqnarray}
&&Var {\hat x}_c =\frac{1}{4} + \frac{1}{4}|\mu  \pm \nu|^2 \nonumber \\
&&Var {\hat x}_s =\frac{1}{4} + \frac{1}{4}|\mu \mp \nu|^2
\end{eqnarray}

Then the total minimum bound is 
\begin{equation}
\min_{\mu, \nu} (Var {\hat x}_c + Var {\hat x}_c )= 1, \quad \mu=1, \nu=0
\end{equation}
This is equivarent to Yuen-Lax bound.\\

\subsubsection{\textbf{Importance of signal to noise ratio in estimation}}
In general, one should keep the following in mind. That is, the performance of the estimation cannot be 
evaluated only by the estimation bound. 
Following the communication theory, one needs the concept of the signal to noise ratio 
which means the operational meaning in the technology.

Let us show an example of the evaluation for continuous parameter.
The squeezed state reduces the quantum noise of one of the quadrature amplitudes, 
and the purpose is to utilize the signal with that low noise.
However, we have the following formulation as follows:\\

$\textbf {Theorem 7}$$\{Yuen\}$[29] : \\
We assume that $\rho =|\beta, \mu_s, \nu_s ><\beta, \mu_s, \nu_s|$ as a signal state.
Let us define the SNR of this state as follows:
\begin{equation}
SNR=\frac{(Tr \rho X_c)^2}{Tr \rho (X_c - <X_c>)^2}
\end{equation}
Under the constraint of the signal energy $Tr \rho {\bf a}^{\dagger} {\bf a} \le N_s$, the maximum $SNR$ is given by
\begin{equation}
SNR_{max}=4 N_s(N_s+1)
\end{equation}
where 
\begin{equation}
\mu_s = \frac{N_s+1}{\sqrt {2N_s+1}}, \quad \nu_s = \frac{N_s}{\sqrt {2N_s+1}}
\end{equation}
The estimation bound is given by
\begin{equation}
Var {\hat x}_c = \frac{1}{4}|\mu_s - \nu_s|^2=\frac{1}{4}\times \frac{1}{2N_s +1}
\end{equation}
When the state is coherent state, $SNR=4N_s$\\

As mentioned above, low noise does not mean good performance. $SNR$ is the final evaluation for communication technology.

When there is an energy loss $\epsilon$ in channel, the SNR is 
\begin{equation}
SNR=\frac{4\epsilon N_s(N_s+1)}{\epsilon+(1-\epsilon)(2N_s+1)}
\end{equation}
The estimation bound is 
\begin{equation}
Var{\hat x}_c=\frac{\epsilon}{4}\times \frac{1}{2N_s+1} + \frac{1-\epsilon}{4}
\end{equation}
When the loss is large, the above SNR becomes $SNR=2N_s$, which is worse than the coherent state.

Since evaluations for problems in technology require the operational meaning, 
we emphasize the points to note in 
\textbf{``communication theory and mathematical statistics"}(Fig.3).\\

\begin{figure}
\centering{\includegraphics[width=8cm]{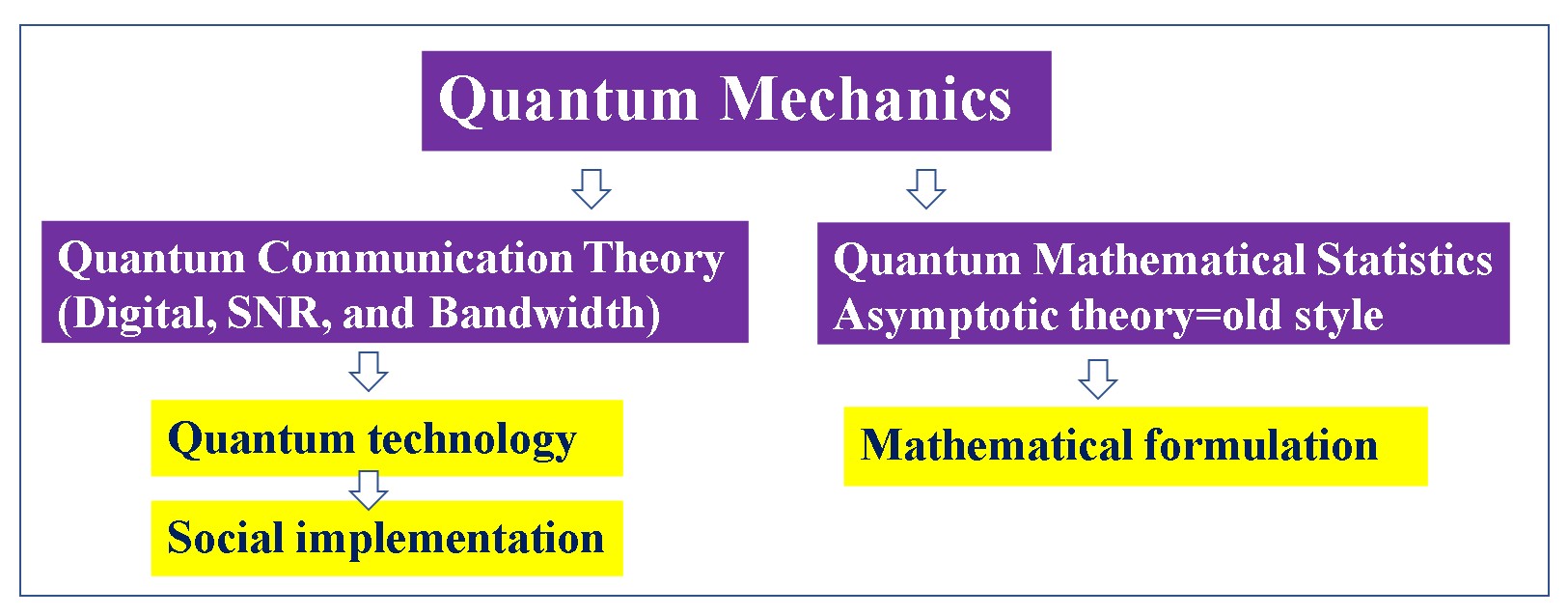}}
\caption{The fundamental concepts of quantum communication theory and quantum mathematical statistics are different, 
though both have certain similarities.
The former must ensure technical superiority in conjunction with parameters such as communication speed (bit/sec), SNR, and bandwidth.
The latter provides a formulation for statistics, 
with asymptotic properties etc. as its main objective. Thus, it does not provide the operational meaning. 
The relationship between the two is the same in classical theory.}
\end{figure}

\section{\textbf{Quantum Shannon information transmission theory}}
Shannon theory is a design theory for transmitting signals through noisy communication channels with 
as few errors as possible. Shannon defined mutual information as the evaluation function, and its maximum value 
is called the channel capacity. Its operational meaning is the maximum rate(a kind of efficiency) at which errors can be minimized.

In this section, we explain the foundation of design theory for systems in which communication signals are transmitted by 
optical signal with quantum nature. 
From Theorem 1, it is necessary to assume a coherent state for quantum design to provide advantages over 
existing communication capabilities in real-world environments.

\subsection{\textbf{Channel Capacity}}
\subsubsection{\textbf{Finite discrete alphabet system}}

The Shannon mutual information in the quantum measurement channel is written as follows:

\begin{equation} 
{J}_{\rm 1}(\mit \xi,\Pi )=\sum\nolimits\limits_{\mit i\rm =1}^{\mit M} 
\sum\nolimits\limits_{\mit j\rm =1}^{\mit M'} 
{\xi }_{\mit i}{\mit P}\rm (\mit j\rm |\mit i\rm ) 
\ln\left[{{\mit P\rm (\mit j\rm |\mit i\rm ) 
\over \sum\nolimits\limits_ {\mit k}^{} {\xi }_{\mit k}
{\mit P}\rm (\mit j\rm |\mit k\rm )}}\right]
\end{equation}
where $P(j|i) =Tr \rho_i \Pi_j$.
The theory of maximaization of mutual information is twinned with the above quantum decision theory 
and it is denoted as follows:
\begin{equation} {C}_{\rm 1}= \sup_{\mit \xi,\Pi} {J}_{ \rm 1}(\mit \xi,\Pi ) 
\end{equation}
 Thus, it has the same theoretical structure as the optimal theory for the detection theory. 
 As a result, the basic result was derived by Holevo [2]:\\

$\textbf {Theorem 8}$$\{Holevo\}$\\
The necessary condition for maximum mutual information with respect to the decision operators for a simple set of states 
is given as follows:
\begin{eqnarray}
&&P(j|i) =Tr \rho_i \Pi_j \nonumber  \\
&& {\bf F}_j = \sum_{i} \xi_i \rho_i \log  \{\frac {P(j|i)}{\sum_l \xi_l P(j|l)}\} \nonumber \\
&& \Pi_j[{\bf F}_j - {\bf F}_i]\Pi_i =0,  \forall i,j 
\end{eqnarray}
where the solution is called accessible information.\\

On the other hand, there is a problem of Davies in the case of the optimization of mutual information [30].
A partial solution to this is given by [31] under Jozsa's guidance.\\

We consider the concrete property of the above. Several properties of the mutual information were discussed 
in the serial papers of  M.Osaki [32]. As in signal detection theory, the mutual information obtained 
by the optimal decision operator is greater than that obtained by a classical measurement process.
The most useful results are as follows:\\

$\textbf {Remark 3}$\\
When the signal states are the group covariant, the quantum Bayes and minimax decision operators satisfy 
the above necessary condition for the mutual information.
In a real environment with large losses or background noise, no quantum state exists that can exceed 
the mutual information provided by communication based on coherent states.\\

Now we still have a difficult problem that  is a proof of the sufficiency.
According to Osaki's analysis, the quantum minimax decision operator may provide 
the maximum mutual information in the practical region such as $|\alpha|^2 \gg 1$ in the set of coherent states.

On the other hand, let us consider $n$-th extension of the alphabet and adopt the decision operator 
based on entangled measurement. Then the following conjecture was imposed.
\begin{equation}
C_1 \le C_2 \le \dots \le C_n \dots \le C_{\infty} = \lim_{n \rightarrow \infty} \frac{1}{n} C_n
\end{equation}
For the pure state channel, Hausladen et al [4] proved it and  Holevo, Schumacher and Westmoreland  proved it for 
the general states [5,6].\\

$\textbf {Theorem 9}$\\
 $\{Holevo\cdot Schumacher \cdot Westmoreland \}$\\
The upper bound of maximum mutual information and the capacity are given by the Holevo information as follows:
\begin{eqnarray}
&&C_1 \le S( \rho_T) - \sum_{i=1}^{M} \xi_i S(\rho_i)=I_H\\
&&where \quad \rho_T = \sum_{k=i}^M \xi_i \rho_i, \nonumber\\
&&Then \quad C_{\infty}=C_{Holevo}=\max_{\xi} I_H
\end{eqnarray}
where $S(\rho)=-Tr \rho \log \rho $ is the von Neumann entropy.\\

The above theorm is the channel capacity formula for quantum Shannon information transmission system.\\

$\textbf {Remark 4}$\\
The maximum absolute value of $Holevo \cdot Schumacher \cdot Westmoreland$ (HSW) capacity in a real environment is 
given by the coherent state.\\

The numerical properties of HSW capacity for coherent state signals have been analyzed by our group  [33,34].\\

\subsubsection{\textbf{Infinite alphabet system (continuous)}}

After the analysis for discret alphabet system, it was extended to analog signals with constrained power inputs among 
which the channels with additive quantum Gaussian noise. This means the issue of infinite alphabet scheme.
Here we denote a simple result on the final capacity formula of quantum Gaussian channel of free space.
According to quantum Shannon theory established by Holevo and others, the capacity formula of optical 
quantum communication for free space lossy Gaussian channel is discussed in [35, 36, 37]:\\

${\bf Theorem 10}$:$\{Holevo \cdot Sohma \cdot Hirota\}$\\
The capacity formula of the single mode quantum lossy Gaussian noise channel is given as follows [36]:
\begin{eqnarray}
C_{Holevo}&=&g(N_s+N_{Th})-g(N_{Th})\nonumber \\
&=&(N_s + N_{Th} +1)\log(N_s +N_{Th} +1)\nonumber \\
&&-(N_s + N_{Th})\log (N_s + N_{Th}) \nonumber \\
&&-(N_{Th} +1)\log (N_{Th} +1) \nonumber \\
&&+ N_{Th}\log N_{Th}
\end{eqnarray} 
where the above is given by coherent state, and where $g(x)=(x+1)\log (x+1) -x\log x$, 
$N_s$ and $N_{Th}$ are average photon numbers of signal and additive noise at receiver, 
and $1$ of $(1+N_{Th})$ means the quantum noise, 
respectively.
\\

The above formula is in general greater than the Shannon classical capacity.
\begin{equation}
C_{Shannon}=\log (1 +\frac{N_s}{1+N_{Th}})
\end{equation}
where the above formula is given by heterodyne receiver. 
Thus, the main parameter of capacity in Shannon theory for continuous alphabet is the $SNR$.\\

\subsubsection{\textbf{Quantum advantage}}
Let us discuss the quantum advanatge. 
In the transmission system of Shannon information, the quantum advantage for discret alphabet is given by
\begin{equation}
C(classical) \le C_1 \le C_H
\end{equation}
where $C(classical)$ means the capacity for digital optical fiber communication system designed by 
conventional communication theory. 
The above relation of quantum advantages results from the error mitigation effect 
in the quantum decision mechanism and coding 
designed by quantum detection theory.

On the other hand, the quantum advantage in the case of infinite alphabet (continuous) is given by 
\begin{equation}
C_{Holevo} - C_{Shannon} \ge 0
\end{equation} 
When the back ground noise is very large, the quantum advantage disappers.\\

\subsubsection{\textbf{Implementation problem}}
It is important to discuss technologies that will actually realize quantum gain in terms of capacity.
The quantum gain for $C_1$ comes from the effect of decision operator of single shot. 
The quantum gain of $C_H$ comes from both of 
decision operator based on entangled measurment and coding scheme.
Below are some examples of discrete and continuous systems.\\
\\

\textbf{(a)} Finite discrete case: 

The problem in this case is a issue of super additivity of the capacity $C_n$ with respect to $n$.
There are two elements to achieving superadditivity. One is a code construction consisting of a sequence of quantum states.
The second is a construction of decision operator and its generalization based on entangled measurement : Eq(12). 
Many pioneering analyses of the former have already been published, but the latter remains a very difficult problem.
Several challenging attempts based on measurement effect have been made [33, 39, 40, 41, 42], 
and some analytical examples based on coding and measurement have been given by [43, 44].\\

\textbf{(b)} Continuous case:
 
Generally, quantum gain disappears when the background noise is large or the signal energy is large. 
Therefore, we discussed how to achieve this when there is no background noise and the received signal is very weak 
from an ultra-long distance. In this case, it is known that the capacity can be attained by the discretization [45].
\\

\subsection{\textbf{Quantum Reliability Function and Quantum Cut-off Rate}}
\subsubsection{\textbf{Reliability function}}
The operational meaning of Shannon mutual information is the efficiency of coding and decoding in a noisy channel. 
The unit is bits/symbol and it is not the actual amount of information.
Reliability functions are introduced to clarify the operational meaning of Shannon's mutual information and 
channel capacity. Here, we show their quantum equivalents. In general, the reliability function is defined as follows:
\begin{equation} 
E_{Q'}\rm (\mit R\rm )=\lim_{\mit n\rm \rightarrow \infty }
sup {1 \over \mit n}\rm \ln{1 \over {\mit P}_{e}}
\end{equation}
The upper bound of the average error probability of a code of length $n$ is as follows:
\begin{equation} 
{P}_{e}\rm \le {\mit e}^{\rm -\mit n{E_{Q'}}^{}\rm (\mit R\rm )}
\end{equation}
We will avoid mathematical details and discuss in a simplified form.\\

$\textbf {Theorem 11}$ $\{Burnashev \cdot Holevo \}$ [8,9]\\
The upper bound on the average error probability of a pure state code for a quantum channel is given by:
\begin{eqnarray} 
P_{e} &\le& 2\exp \{-\max_{\xi}\max_{0 \le s \le 1}n[\mu_Q(\rho_{\xi}, s) -sR]\}\nonumber \\
&\equiv& exp[-n\textbf{E}_Q(R)]
\end{eqnarray}
where 
\begin{eqnarray}
\textbf{E}_Q(R)&=&\max_{\xi}\max_{0 \le s \le 1}n[\mu_Q(\rho_{\xi}, s) -sR]\\
\mu_Q(\rho_{\xi}, s)&=&-\ln[Tr \rho_{\xi}^{1+s}], \\
 \rho_{\xi}&=&\sum_{j=1}^M \xi_j |\psi_j ><\psi_j|\nonumber
\end{eqnarray}
where $M$ is the number of symbol.\\

Now the maximization problem becomes:
\begin{equation} 
{\rm \partial \over \partial \mit s}\rm 
\left[{{\mit \mu }^{}\rm (\mit s\rm ,\mit \xi \rm )- 
\mit sR}\right]\rm ={\partial \left[{{\mit \mu }^{}\rm 
(\mit s\rm ,\mit \xi \rm )}\right] \over \partial \mit s}
\rm -\mit R\rm =0 
\end{equation}

\begin{equation} 
{\rm \partial \mit \mu \rm (\mit s\rm ,\mit \xi \rm ) 
\over \partial \mit s}\rm = {-Tr{\rho}_{\xi }^{\rm 1+\mit s}
\rm \ln{\rho}_{\xi } \over \rm Tr{\rho}_ {\xi }^{\rm 1+\mit s}}
\rm ={-\sum\nolimits\limits_{}^{} {\mit \lambda }_ {\mit j}^{\rm 1+\mit s}
\rm \ln{\mit \lambda }_ {\mit j} \over \rm \sum\nolimits\limits_{}^{} 
{\mit \lambda }_ {\mit j}^{\rm 1+\mit s}} 
\end{equation}
where $\lambda$ is an eigenvalue of $\rho_{\xi}$ and it is given by eigenvalue of the following matrix:
\begin{equation} 
\left[ \begin{array}{ccc} \sqrt{\xi_{1}\xi_{1}}\langle\psi_{1}
\vert\psi_{1}\rangle & \cdots & \sqrt{\xi_{1}\xi_{M}}
\langle\psi_{1}\vert\psi_{M}\rangle \\ \vdots & \ddots & 
\vdots \\ \sqrt{\xi_{M}\xi_{1}}\langle\psi_{M}
\vert\psi_{1}\rangle & \cdots & \sqrt{\xi_{M}\xi_{M}}
\langle\psi_{M}\vert\psi_{M}\rangle \end{array} \right] 
\end{equation} 
The above formula assumes decision operator based on entangled measurement.

If quantum measurements are individual measurements, the reliability function is given by
\begin{equation} 
{\bf E}_{semi}=-\mit s R-\ln\sum_{j=1}^{M'} 
\left(\sum_{i=1}^M \xi_i{P\left(j|i\right)}^{1/1+\mit s}\right)^ {1+\mit s}
\end{equation} 
where $P(j|i)=Tr \rho_i \Pi_j$ and the optimum $\{\Pi\}$ is given by the minimum error probability conditions [10,11,12].
So this is called semi quantum reliability function.
Ban and Kurokawa clarified the difference between two definitions [13,14].\\

\subsubsection{\textbf{Quantum cut-off rate}}
The quantum cut-off rate is defined from formulation of the quantum reliability function of 
Burnashev $\cdot$ Holevo as follows: \\

$\textbf {Definition 9}$\\
The quantum cut-off rate is defined as follows:
\begin{eqnarray} 
{\bf R}_{\mit Q} & \equiv & \max_{         \left\{{\mit \xi }_{\mit i}\right\}         }         
\left[                 {\mit \mu}\left({\rho_{\xi} }_{\mit i},                 
{\mit s}{\rm =1}\right)         
\right] \nonumber\\ & = & \max_{\left\{{\mit \xi}_{\mit i}\right\}} 
\left[{\rm -\ln}\sum\nolimits\limits_{\mit i\rm =1}^{\mit M}  
\sum\nolimits\limits_{\mit j\rm =1}^{\mit M}  
{\mit \xi}_{\mit i}{\mit \xi }_{\mit j} 
{|\langle{\psi }_{\mit i} | {\psi }_{\mit j}
\rangle|}^{\rm 2} \right]\nonumber\\ & = & 
-{\rm ln}\left[{\min_{\left\{{{\mit \xi }_{\mit i}}
\right\}} \sum\nolimits\limits_{\mit i\rm =1}^{\mit M}
\sum\nolimits \limits_{\mit j\rm =1}^{\mit M}  
{\left({\mit \Gamma}\right)}_{\mit i\mit j} 
{\xi }_{\mit i}{\mit \xi }_{\mit j}}\right]
\end{eqnarray} 
where $(\Gamma)_{ij}=|<\psi_i|\psi_j>|^2$.\\
 
When the optimaization w.r.t $\xi$ is done, we have the following simple form.\\

$\textbf {Theorem 12}$ $\{Ban \cdot Kurokawa \cdot Hirota \}[13,14]$\\
The quantum cut off rate is given by 
\begin{equation} 
{\bf R}_{\mit Q}\rm = -ln\left[{\sum\nolimits\limits_{\it i\rm =1}^{\it M}  
\rm \sum\nolimits\limits_{\it j\rm =1}^{\it M}  
{\left({{\mit \Gamma}^{\rm -1}}\right)}_{\it i\it j}}\right]. 
\label{eqn:RQ-2} 
\end{equation} 
under the condition for the optimum values of $\xi$ as follows:
\begin{equation} 
{\tilde{\it \xi }}_{\it i}= \frac {{\sum\nolimits\limits}_{\it j\rm =1}^{\it M}  
{\left({\mit \Gamma}^{\rm -1}\right)}_{\it i\it j}} 
{{\rm \sum\nolimits\limits}_{\it i\rm =1}^{\it M}  
{\sum\nolimits\limits}_{\it j\rm =1}^{\it M}  
{\left({\mit \Gamma}^{\rm -1}\right)}_{\it i\it j}}> 0  \quad  \forall i
\end{equation} 

When a priori probability is uniform, it bexcomes as follows:
\begin{eqnarray} 
{\bf R}_{Q} & = & \ln\left[\frac {M}{\sum_{j=1}^{s}
\vert\langle\psi_{i}\vert \psi_{j}\rangle\vert^{2}}\right]\nonumber\\ 
& = & \ln\left[\frac {M}{\langle\psi
\vert \hat{\mit\Phi}\vert\psi\rangle}\right].
\end{eqnarray} 
where $\hat{\mit\Phi}=\sum_{i=1}^M |\psi_i ><\psi_i |$.\\

Thus, the relation between the reliability function and cut off rate is given by 
\begin{equation} 
\textstyle{ {P}_{e}\le2{\rm \exp}\left[{\rm -\mit n{\bf E}_{Q}
\rm (\mit R\rm )}\right] \le 2{\rm \exp}
\left[{\rm -\mit n(\bf R}_Q-R)\right]} 
\end{equation}

Here let us introduce the definition of Helstrom based on individual measurement.\\

$\textbf {Definition 10}$\\
Here a cutoff rate based on quantum individual measurement is defined as follows:
\begin{equation} 
{\bf R}_{semi} = \max_{         
\left\{{\mit \xi }_{\mit i}\right\}         }         
\max_{         \left\{{\hat {\mit \Pi}}_{\mit i}\right\}         }         
\left({-\rm ln}\sum\nolimits \limits_{\mit j\rm =1}^{\mit M'}  
\left[ \sum\nolimits \limits_{\mit i\rm =1}^{\mit M} 
{\mit \xi }_{\mit i}\sqrt{\left[{\rm Tr} {\hat \rho}_{\mit i}
{\hat {\mit \Pi}}_{\mit j}\right]} \right] ^2 \right) \label{eqn:rb1} 
\end{equation} 
And the upper bound of the above formula is given by
\begin{equation} 
{\tilde {\bf R}}_{semi}= \max_{\left\{{\mit \xi }_{\mit i}\right\}} 
\left[ -{\rm ln} \sum\nolimits\limits_{\mit i\rm =1}^{\mit M} 
\sum\nolimits\limits_{\mit j\rm =1}^{\mit M}  
\left({\mit G}\right)_{\mit i\mit j} {\mit \xi}_{\mit i}
{\mit \xi}_{\mit j}\right] \label{eqn:rb2} 
\end{equation} 
where $G_{i,j}=|<\psi_i |\psi_j >|$. As a result, we have 
\begin{equation}
{\bf R}_{semi} \le {\bf R}_Q \le {\bf E}_Q
\end{equation}
The example will be given in the next section.

In the above, we introduced the special case such as pure state. For general case, refer Holevo's paper [9].\\

\begin{figure}
\centering{\includegraphics[width=8cm]{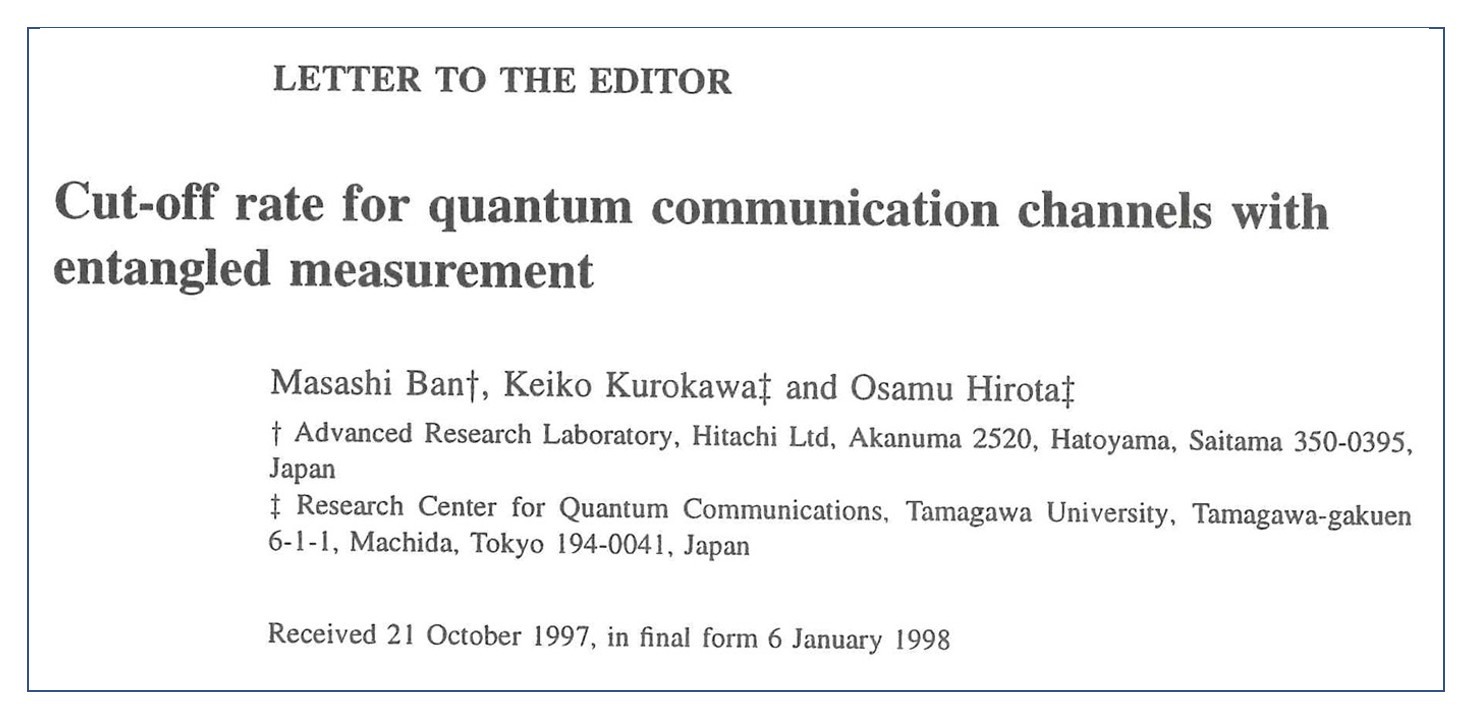}}
\caption{Effect for the decision operator based on entangled mesurement to reliablity function and cut off rate. 
Quantum Semiclass. Opt. vol-10, L7-L12, 1988. }
\end{figure}

\section{\textbf{Examples of reliability function and cut-off rate}}
\subsection{\textbf{Finite Discrete Alphabet System}}
Let us show some examples of the reliability function and cut-off rate for the finite alphabet. \\

\subsubsection{\textbf{Analytical method}}
The signal system is set to 3-ary PSK, which is composed of coherent states 
that enable ultra-high speeds over long distances.
\begin{equation}
\left|{{\psi }_{1}}\right\rangle\rm=\left|{\alpha }\right\rangle\rm,  
\left|{{\psi }_{2}}\right\rangle\rm=\left|{\mit \alpha {e}^{i{\rm 2 \over 3}
\mit \pi }}\right\rangle\rm ,
\left|{{\psi }_{3}}\right\rangle\rm=\left|{\mit \alpha {e}^{\rm -\mit i
{\rm 2 \over 3}\mit \pi }}\right\rangle.
\label{eqn:3PSK}
\end{equation}
In this case, the modified Gram matrix is as follows:
\begin{equation}
\frac{1}{3}\left[
\begin{array}{ccc}
1&
K_c+{\rm i}K_s&
K_c-{\rm i}K_s\\
K_c-{\rm i}K_s&
1&
K_c+{\rm i}K_s\\
K_c+{\rm i}K_s&
K_c-{\rm i}K_s&
1
\end{array}
\right],
\label{eqn:G2}
\end{equation}
The eigenvalues of the above are
\begin{eqnarray}
\lambda_1& = &\frac{1+2K_c}{3},\nonumber\\
\lambda_2& = &\frac{1-K_c-\sqrt{3}K_s}{3},\nonumber\\
\lambda_3& = &\frac{1-K_c+\sqrt{3}K_s}{3}
\end{eqnarray}
where
\begin{eqnarray}
K_c& = &{\rm exp}\left[-\frac{3N_s}{2}\right]\cos
\left[\frac{\sqrt{3}N_s}{2}\right],
\nonumber\\
K_s& = &{\rm exp}\left[-\frac{3N_s}{2}\right]\sin
\left[\frac{\sqrt{3}N_s}{2}\right]
\end{eqnarray}
where $N_s$ is the average photon number per pulse at the receiver.
From these, we can derive the quantum reliability function and quantum cutoff.

On the other hand, if we assume individual measurements, we will adopt individual optimal decisions. 
In that case, the elements of the communication channel will be as follows:

\begin{eqnarray}
P(1|1)& = &P(2|2)=P(3|3)\nonumber\\
& = &\frac{1}{9}\{3+2\sqrt{\beta_1\beta_2}+\nonumber\\
& &2\sqrt{\beta_2\beta_3}+
2\sqrt{\beta_3\beta_1}\},\nonumber\\
P(1|2)& = &P(2|3)=P(3|1),\nonumber\\
P(1|3)& = &P(2|1)=P(3|2),\nonumber\\
& = &\frac{1}{9}\{3-\sqrt{\beta_1\beta_2}-\nonumber\\
& &\sqrt{\beta_2\beta_3}-
\sqrt{\beta_3\beta_1}\}
\end{eqnarray}
where
\begin{eqnarray}
\beta_1& = &1+2K_c,\nonumber\\
\beta_2& = &1-K_c-\sqrt{3}K_s,\nonumber\\
\beta_3& = &1-K_c+\sqrt{3}K_s
\label{eqn:lam123}
\end{eqnarray}
Fig.5 shows the numerical example. The communication system we set up here is intended for deep space communication, 
where the transmitted signal is sufficiently large and the transmission attenuation is extremely large, 
resulting in an extremely small received signal.\\

\begin{figure}
\centering{\includegraphics[width=8cm]{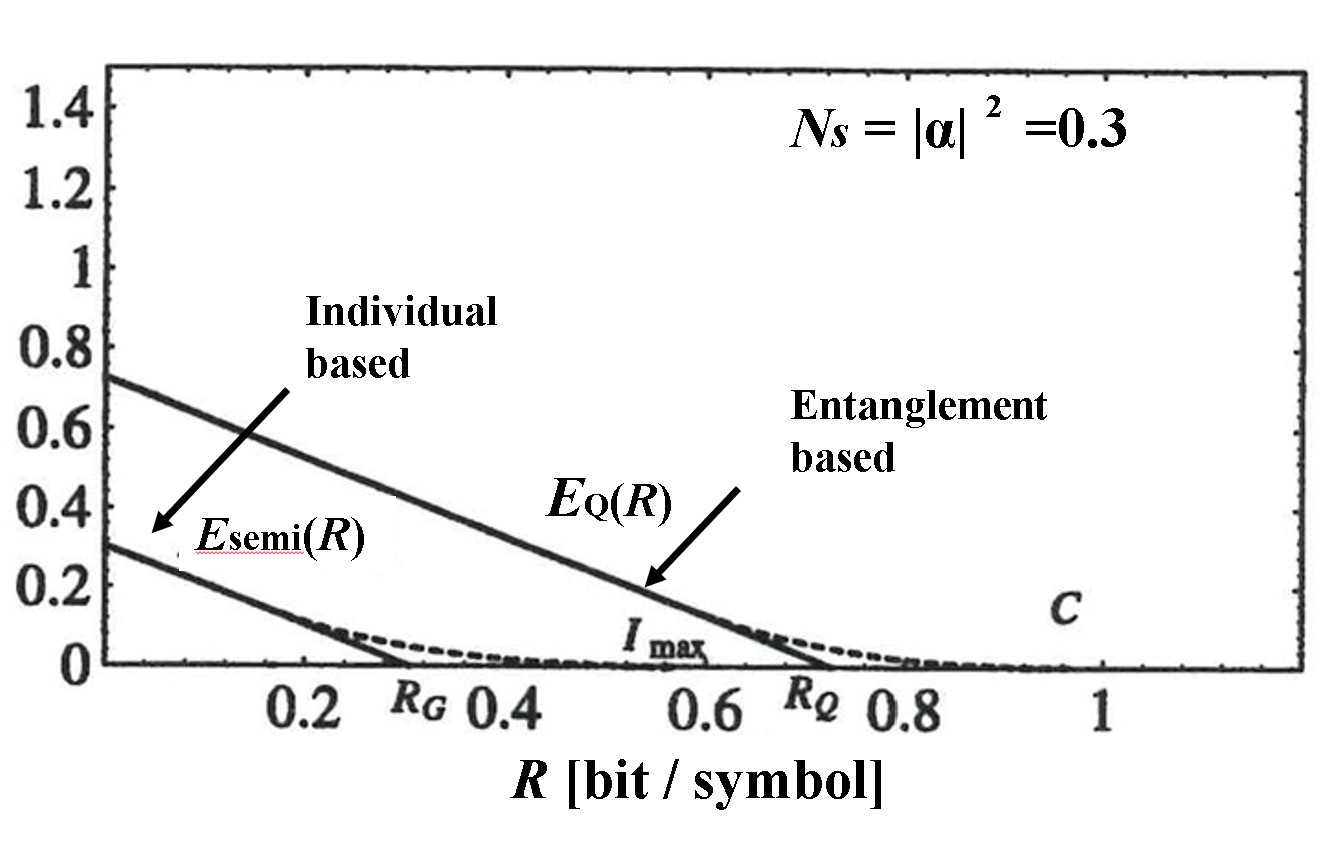}}
\caption{Numerical examples of reliabilty function and cut off rate for the decision operators based on 
entangled measurement and indivisual measurement.[38]}
\end{figure}

\begin{figure}
\centering{\includegraphics[width=7cm]{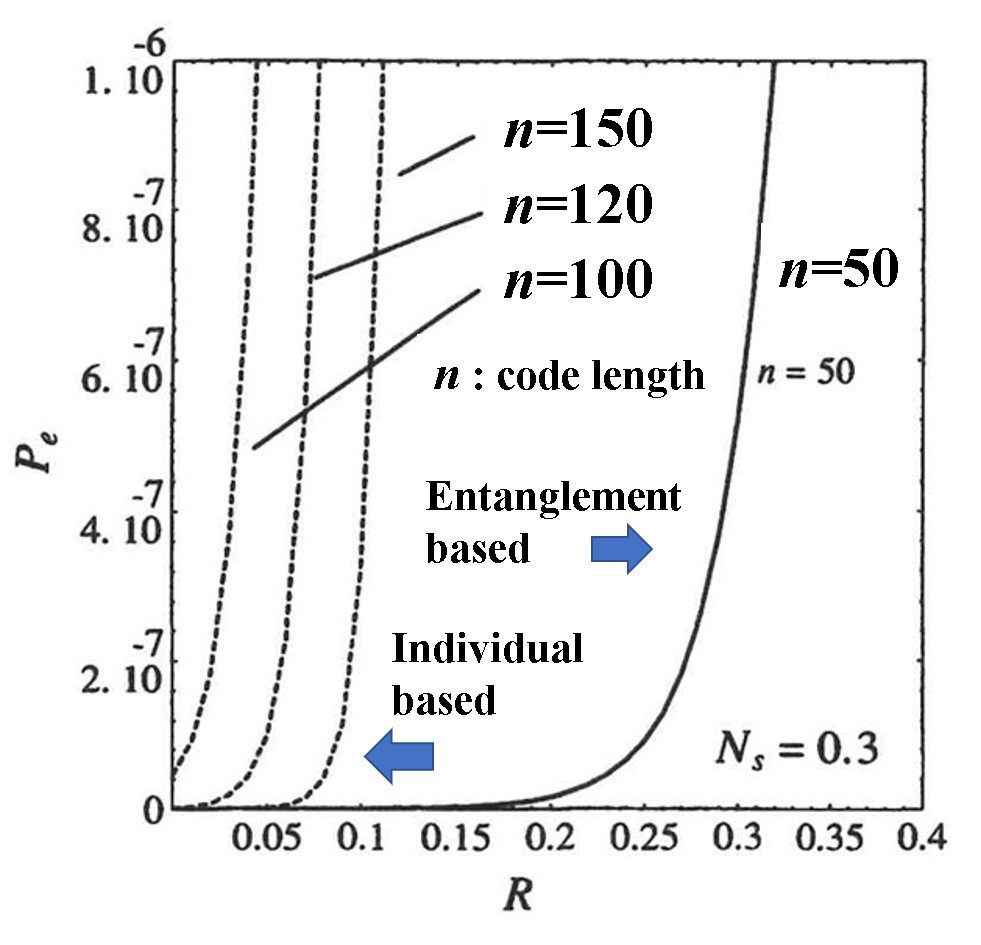}}
\caption{Numerical examples of the quantum advantage to code length by decision operators based on 
entangled measurement and indivisual measurement.[38]}
\end{figure}

\subsubsection{\textbf{Quantum advantage of decision operator based on entangled measurement}}
The role of the quantum reliability function theory is to show how small the error rate of the transmitted code can be 
when the transmission rate of code is set. 
In the design theory of conventional optical communications, classical devices such as heterodynes and 
energy detectors are used as receivers, 
and reliability functions are evaluated using classical theory.
The question is what benefits the quantum theory introduced here will bring to conventional optical communications.
In quantum theory, quantum effects appear in quantum measurements, i.e., at the receiver.
From the above theory, individual measurements only reduce the effect of quantum noise on individual signals. 
On the other hand, entangled measurements can take into account the quantum interference effect among signals due to measurement
in addition to the individual quantum effects, thereby further reducing the quantum noise effect and 
increasing the reliability function and cutoff rate.
As a result, when the code error rate is fixed (depending on the application), 
the code length required to achieve the same error rate can be significantly shortened. See Fig.6.\\

\subsection{\textbf{Infinite Alphabet System (continuous)}}
\subsubsection{\textbf{Reliability function}}
Here we introduce the case of continuous systems with an infinite alphabet [46]. 
This discussion is extremely important because it can clarify how quantum gain appears, i.e., in comparison 
with discrete systems.

Let us take the input alphabet $A$ an arbtrary Borel subset in a finite dimensional Euclidean space.
The input is described by an an a priori probablity $\xi(x)$ on $A$.
The energy constraint is posed as follows:
\begin{equation}
\int_A (x-<x>)^2 \xi(x) dx \le N_s
\end{equation}

We adopt the product Hilbert space $H^{\otimes n}=H\otimes H \otimes \dots \otimes H$ with the input alphabet $^n$ 
consisting of code words $W=(x_1, x_2, \dots, x_n)$ of length $n$, and the density operator
\begin{equation}
\rho_W=\rho (x_1) \otimes \dots \rho (x_2) \dots \otimes \rho (x_n)
\end{equation}
We define the set of code of size $M$ that is a sequence $\{W(1),W(2), \dots,W(M)\}$.
The $\{\Pi^c_j\}$ is a family of decision operatots in $H^{\otimes n}$ based on entangled measurement, 
satisfying $\sum_{j=1}^{M} \Pi^c_j =I$.
The error probability for a codeword is $P(k|j)=Tr \rho_{W(j)} \Pi^c_k$ where $j, k=1,2,\dots, M$.

Let us consider the speed of the exponential decay of the error probability when $n \rightarrow \infty$ and $M=e^{nR}$ 
below the following capacity: 
\begin{equation}
C_H=\max_{{\cal P}(x)}[\hat{\mit H}(\Lambda [\rho_T(\xi(x))]
-\int \hat{\mit H}(\Lambda [\rho(x)])\xi(x)dx
\end{equation}

Following the Shannon idea of random coding, let us consider the random ensemble of $M$ codewords of the length $n$.\\

$\textbf {Theorem 12}$ $\{Holevo \}$\\
Let us define $P_e(M=e^{nR},n)$ as the minimum error probability w.r.t coding and decision operator.
Here we introduce the reliability function as follows:
\begin{equation}
{\bf E}(R)=\lim_{n\rightarrow \infty} \sup \frac{1}{n} \log{\frac{1}{P_e(e^{nR},n)}}
\end{equation}
The lower bound of the reliability function is given by
\begin{equation}
{\bf E}(R) \ge \max \{{\bf E}_{Q_r}(R), {\bf E}_{Q_{ex}}(R) \}
\end{equation}
where $Q_r$ and $Q_{ex}$ mean the random coding bound and expurgated bound, resepectively. They are 
\begin{eqnarray}
{\bf E}_{Q_r}(R)&=&\max_{0 \le s \le 1}(\max_{0 \le p}\max_{\xi} \mu(\xi, s, p) -sR) \\
{\bf E}_{Q_{ex}}(R)&=&\max_{1 \le s}(\max_{0 \le p}\max_{\xi} {\tilde \mu}(\xi, s, p) -sR) 
\end{eqnarray}
and where 
\begin{eqnarray}
\mu(\xi,s,p)&=&-\log Tr(\int e^{p[f(x)-E]}\rho(x) \xi (x))^{1+s} \nonumber \\
{\tilde \mu}(\xi,s,p)&=&-s\log \int \int e^{p[f(x)+f(y)-2E]} \nonumber \\
&\times& (Tr \sqrt {\rho (x) }\sqrt{\rho (y) })^{1/s}\xi(x)\xi(y)
\end{eqnarray}
where $f$ is a function satisfying the condition for the central limit theorem for random variable $f(x_k)$
\begin{equation}
\int f(x)^2 \xi(x)  < E
\end{equation}
This corresponds to the energy constraint.\\

The above equations are primitive formulae of the quantum reliability function theory [46]\\

\begin{figure}
\centering{\includegraphics[width=8cm]{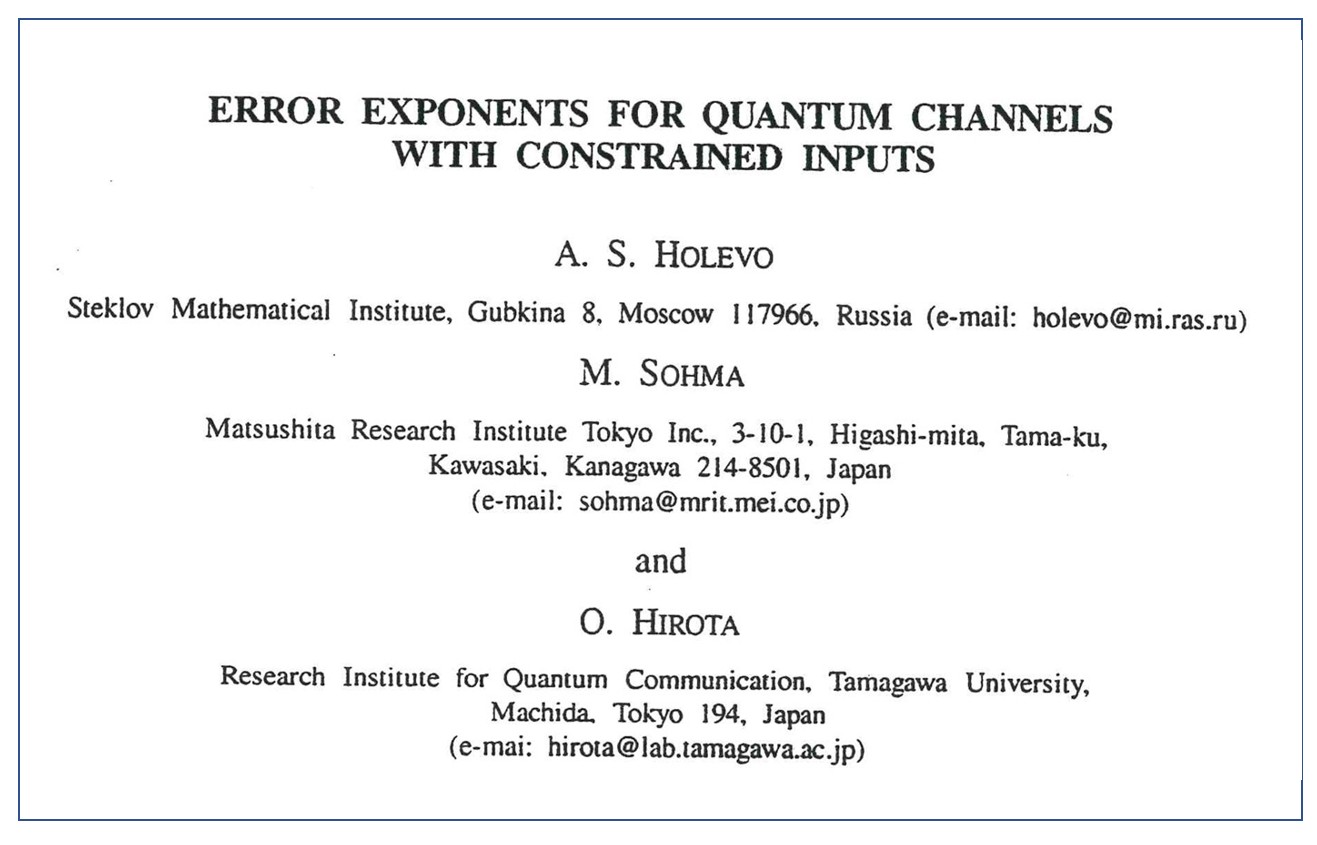}}
\caption{Reliability function and cut off rate theory for infinite alphabet system [46].}
\end{figure}

\subsubsection{\textbf{Cut-off rate}}
In general, it is difficult to obtain the reliability function. 
Therefore, it is useful to define and calculate the cut off rate in the same way as in the discrete system.
It is given by the following relation.
\begin{equation}
{\bf R}_Q =\max_{0 \le p}\max_{\xi} { \mu}(\xi, 1, p) 
\end{equation}
where $s$ is fixed to 1.\\

\subsubsection{\textbf{Example of cut-off rate for Gaussian channel}}
In real communication, the only useful quantum states of continuous quantities are coherent states.
Let us assume that the coherent state alphabet is disturbed by Gaussian noise. 
\begin{equation}
\rho_0=\frac{1}{\lambda +1/2}\sum_{n_p} \big( \frac{\lambda -1/2}{\lambda +1/2}\big)^{n_p} |n_p ><n_p |
\end{equation}
Even in this case, it is difficult to obtain an exact formula for the reliability function.
However we have the exact solutiton of the cut off rate as follows:
\begin{equation}
{\bf R}_Q=2 \big ( \frac{N_{sc}}{2g} + 1 - {\cal D}(N_{sc}/g)\big ) + \log {\cal D}(N_{sc}/g)
\end{equation}
where ${\cal D}(N_{sc})=(1+\sqrt {N_{sc}^2+1})/2$, $N_{sc}$ is energy of code word, and $g=\lambda g_{1/2}(\lambda)$, 
\begin{equation}
g_{1/2}(t)=\frac{1}{2t} \times \frac{(t+1/2)^{1/2}+(t-1/2)^{1/2}}{(t+1/2)^{1/2}-(t-1/2)^{1/2}}
\end{equation}

In this case, the quantum gain is only the entangled effect of the decision process, 
i.e., the double quantum gain disappears as in the discrete system.\\

\subsection{\textbf{Importance of Cut-off Rate and Quantum Advantage}}
According to reliability function theory, when the rate exceeds the cut-off rate and enters the capacity region, 
an extremely long code length is required to achieve a sufficiently small error rate. 
This means increased communication delays, and has come to be seen as the biggest drawback of modern communications technology.
Therefore, rather than achieving capacity, a coding technique that minimizes code length in the cut-off rate region is essential. 
Based on the theory in this section, it is preferable to utilize the effect of entangled measurement to shorten code length 
rather than increase capacity.
This is because the amount of information that can be transmitted (bit/sec) can be increased by orders of magnitude 
with optical communication. Therefore, there is no longer any need to forcefully achieve channel capacity, 
which is efficiency. Rather, delay is the major issue.\\

\section{\textbf{Discovery of a cipher that breaks the Shannon impossibility theorem }}
In the previous section, applications of quantum effects to the modern optical communications were considered for 
achieving the high performance communication, based on the quantum Shannon information theory.

In this section and the sections that follow, we introduce applications of its theory to 
several technologies such as cipher and sensor.
In the case of cipher, to achieve information-theoretic security for symmetric key cipher, 
it is necessary to realize a mechanism that can hide not only data but also secret keys using quantum noise. 
This can consist of modulation and demodulation technologies based on the above theories (Fig.8). 
In this section, we will introduce the basic concept.\\

\begin{figure}
\centering{\includegraphics[width=6cm]{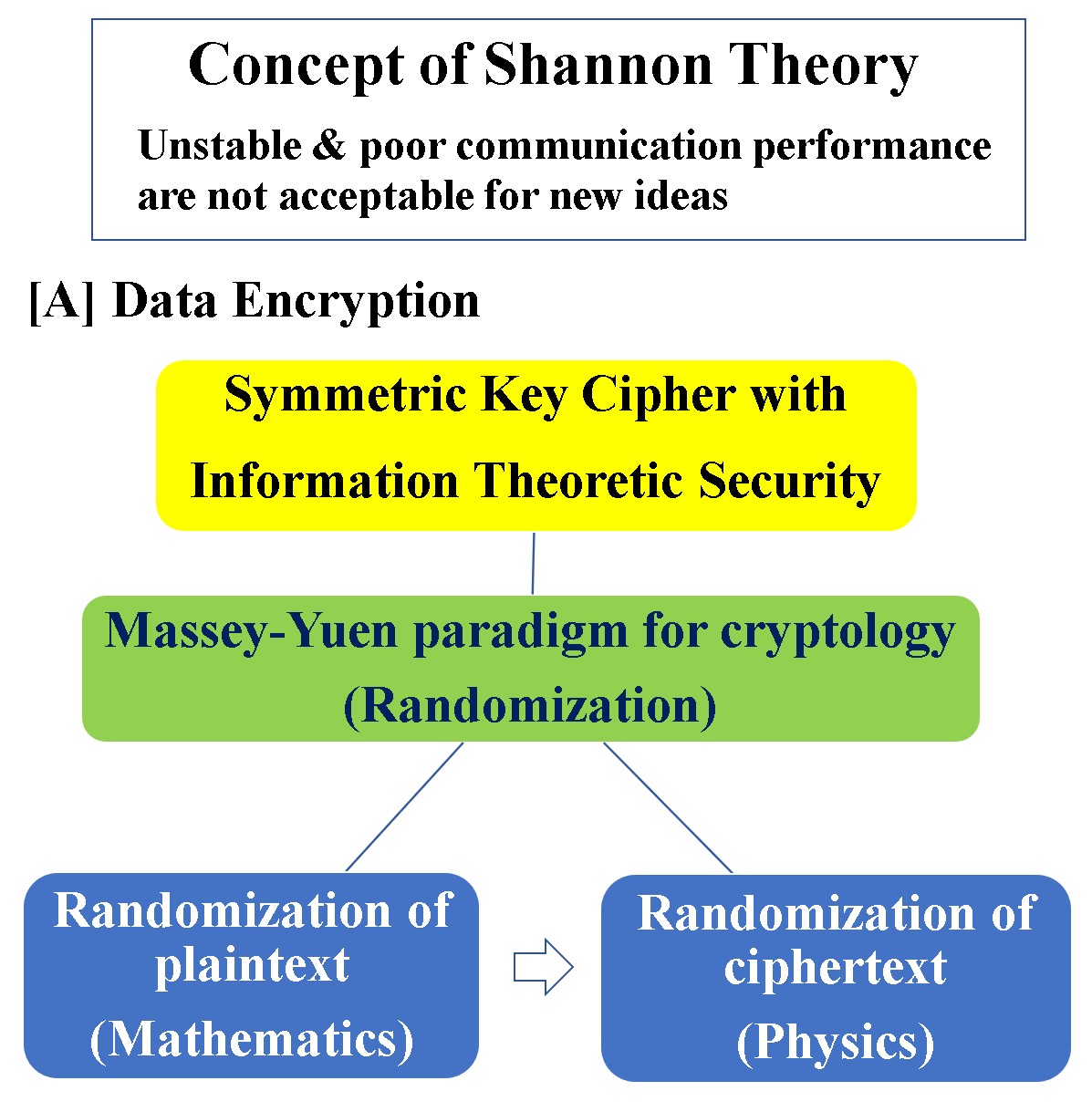}}
\caption{Basic Scenario in the Shannon's concept for cryptology.
One can obtain the drastic quantum advantage in physical encryption based on quantum modulation-demodulation protocol.}
\end{figure}

\subsection{\textbf{New principle for ciphers}}
First of all, we present the method of the application of the quantum Shannon information theory to design of ciphers.
This is the most exciting application of quantum information science, as it promises to completely outperform 
conventional functionality in the real world.

There are several no-go theorems in quantum mechanics.
Quantum Shannon information theory can be considered a theory for designing optimal communication systems that utilize these.

The most important application is to resolve the Shannon impossibility theorem, which limits the security of 
classical cryptography. The Shannon impossibility theorem is given as follows:\\

$\textbf {Theorem 13}$ $\{Shannon \}$\\
The information-theoretic security of symmetric key cipher is limited by the Shannon entropy of the key.
\begin{equation}
H(X|Y) \le H({\rm K}_S)
\end{equation}
where $X$ is the message, ${\rm K}_S$ is the shared secret key, and $Y$ is the ciphertext received by an eavesdropper.
The perfect secure cipher is only given by one time pad cipher. \\

The principle proposed to break the above theorem is  KCQ principle (Keyed communication in quantum noise).
It was disclosed as a white paper in 2000 and made public in 2003 [47].\\

$\textbf {Principle}$ $\{Yuen \}$\\
The non-orthogonality of a set of signal states can be increased by randomization due to a secret key. 
One can creat a differentiation in the reception performances  such that the performance of receiver with key 
to randomized quantum states is superior to that of receiver without key. That is, 
\begin{eqnarray}
{\bar P}^B_e(with-key) &\ll& {\bar P}^E_e(without-key)\\
C^B(with-key) &>& C^E(without-key)
\end{eqnarray}
where ${\bar P}_e$ and $C$ are the error performance of signal detection and the capacity, respectively. 
The index of $B$ and $E$ mean Bob and Eve.
This inequalities are called advantage creation based on KCQ.  \\

\begin{figure}
\centering{\includegraphics[width=8cm]{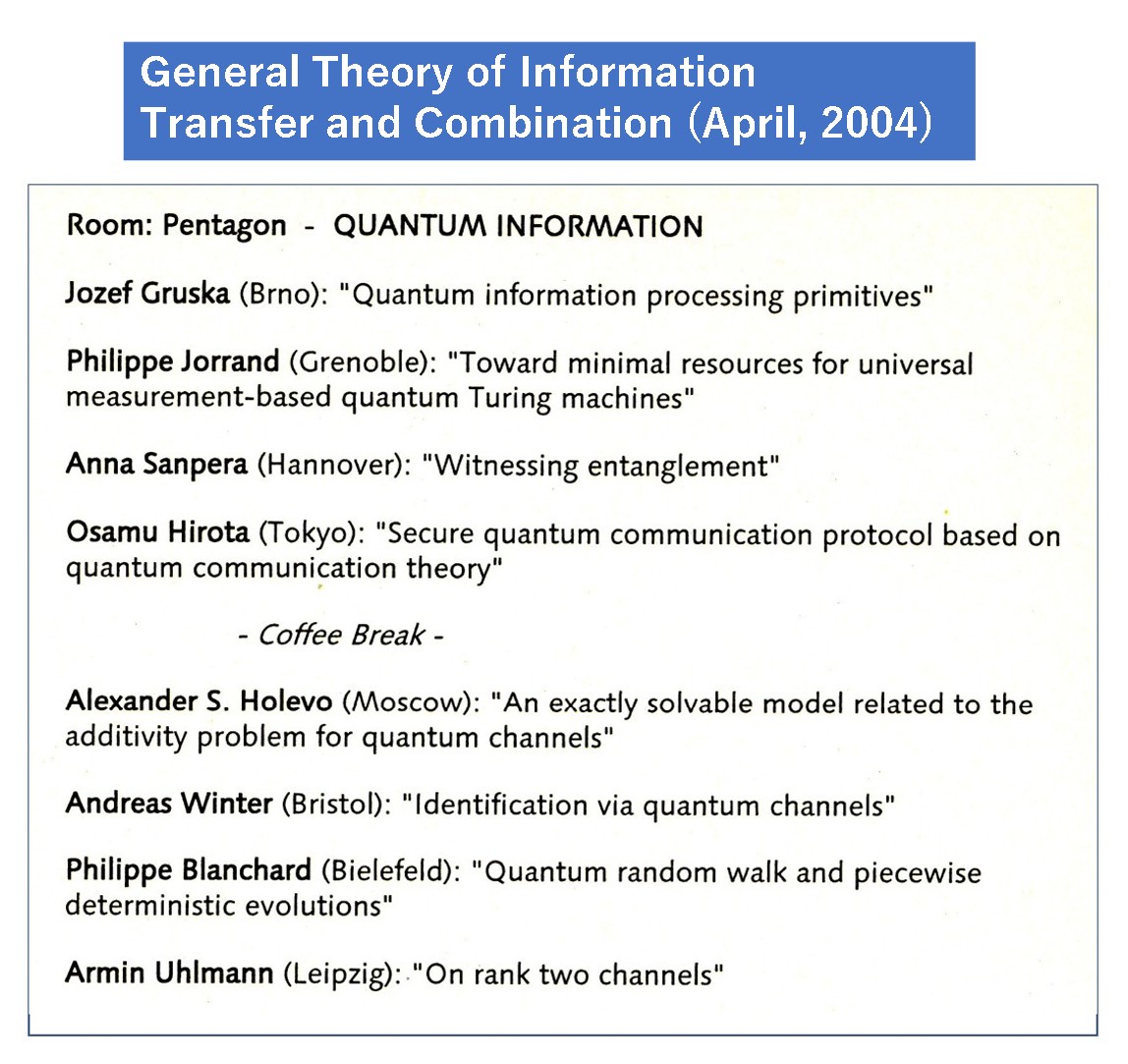}}
\caption{Program of the Conference on General theory of Information Transfer and Combination held at the Center 
for Interdisciplinary Research, Bielefeld University organized by Prof. Rudolf Ahlswede.}
\end{figure}

Yuen's main focus was  quantum key distribution:QKD as an application of own principle.
I introduced the Yuen's concept for lifting the Shannon impossiblity theorem of symmetric key cipher based on 
the above principle in 2004 (Fig.9) [48]. Since then, our collaborative research has brought various important advances 
[49, 50, 51].
In the following, we show a survey how to apply the quantum communication theory to realize the cipher based on KCQ.
Its main concept is to use the quantum modulation as encryption.\\

\subsection{\textbf {Optical quantum modulation as encryption based on the new principle}}
In this section, we will explain the model of encryption-decryption scheme by optical quantum modulation-demodulation, 
and clarify the principle behind the appearance of quantum noise effects to hide the data and the shared secret key.
We mainly deal with quantum stream ciphers and quantum block cipher formats.
\\

(a) \textbf{Quantum stream cipher}

First example of ciphers designed according to this principle is the quantum stream cipher. 
For detailed diagrams of the structure, see the references [49,50,51].

Let us denote the protocol. 
The data (plaintext) is a sequence of binary signals.\\

\textbf{(i)} The sender prepares a big number $(M \gg 1)$ of  communication basis $\{|\alpha_m >, |-\alpha_m >\}$ 
consisting of two non-orthogonal states 
(coherent state with high power) such as
\begin{equation}
\{|\alpha_1>, |-\alpha_1>\},  \dots, \{|\alpha_M>, |-\alpha_M>\}
\end{equation}
where $m=\{1,2,3, \dots, M\}$.
One of them is selcted by using a pseudorandom number generator (PRNG) with a secret key: ${\rm K}_S$. \\
\textbf{(ii)} The sender then transmits binary data by BPSK using 
the selected binary communication basis.
\begin{equation}
0 \quad or \quad 1 \rightarrow  Mapper \rightarrow |\alpha_m> \quad  or \quad |-\alpha_m> 
\end{equation}
where $Mapper$ is the random mapping function due to the same PRNG (See [51]).\\
\textbf{(iii)} A receiver who has the same pseudorandom number with the key can identify the communication basis, 
so he always receives binary signals with small error, because the signal amplitude of binary coherent states is large enough.

A receiver who does not have the key must identify the quantum states of $2M$ that make up the many communication bases, 
which increases the error.
The reason of why is that the discrimination error of a multi-quantum state is larger than that of a binary quantum state 
according to quantum detection theory.

The designer of this cipher could use this error to completely erase the information on the secret key and 
pseudorandom number structure shared by the sender and receiver, optimizing the scheme based on quantum Shannon theory 
(See the reference [51] for the specific protocol and structure).\\

As a result, the ciphertext that Eve receives is completly masked by quantum error, and it becomes possible 
to break the Shannon limit (Shannon impossible theorem), 
which determines the limitation of information-theoretic security of symmetric key cipher. That is, 
\begin{equation}
H(X|Y^{E_q}) \gg H({\rm K}_S)
\end{equation}
where $Y^{E_q}$ is the ciphertext received by the eavesdropper. 

Shannon's focus was only on ciphertext-only attacks, but in modern times it is necessary to deal with known-plaintext attacks. 
For this reason, the following evaluation method has been established [52].

The quantitative evaluation of information-theoretic security is evaluated by the unicity distance for known-plaintext attacks.\\

$\textbf {Definition 11}$:\\
The unicity distance of known plaintext attacks for quantum stream cipher is defined as follows:
\begin{equation}
n^Q_1: H({\rm K}_S|X_{n^Q_1}, Y^{Eq}_{n^Q_1})=0
\end{equation}
It is the minimum value of the ciphertext sequence required for an eavesdropper to be able to guess the key.
Then it is given by the following theorem [52].\\

$\textbf {Theorem 14}$ $\{Yuen \cdot Nair\}$\\
The lower bound of the generalized unicity distance for KPA  is given as follows:
\begin{equation}
n^Q_1 \ge \frac{|{\rm K}_S|}{C^E_1},\quad C^E_1=\max_{\{\Pi^E\}}I(K^R;Y^{E_q})
\end{equation}
where  $C^E_1$ is the maximum amount of the mutual information (Accessible information) by the eavesdropper's measurement 
from set of  quantum state with $K^R$ (running key sequence) as a variable. $\Pi^E$ is the decision operator of Eve.\\

However, the original idea did not guarantee sufficient security against known-plaintext attacks.
In 2007, we solved that problem.
That is, one can attain $C^E_1 \rightarrow 0$ by the generalized randomization [53] (See references [51] for detailed concept).
In other words, while one-time pad requires a secret key the same length as the message, this mechanism can encrypt 
any message in an information-theoretically secure manner using only a short secret key of a few hundred bits.

As a result, these quantum stream ciphers can provide the encryption system of 100 Gbit/sec 
$\sim$ 10 Tbit/sec over a communication distance of 1000 km 
$\sim$ 10,000 Km, guaranteeing information-theoretic security. More details are provided in the subsection :C.\\

(b) \textbf{Quantum block cipher}\\
We can cosider the block cipher form based on the new principle. 
Quantum versions of classical block ciphers are called quantum block ciphers.
Here we introduce the encryption theory for the quantum block cipher using unitary transformation by Sohma [54,55]
 which is also a scheme of ``quantum modulation".

 First, let us consider  $M-$ary coherent state composed by $M$ blocks of coherent state as follows:.
\begin{equation}
|\Phi >=|\alpha_1>|\alpha_2>|\alpha_3> \dots |\alpha_M>
\end{equation}

We consider an operator ${\bf V}$ that extends the Heisenberg commutation relation for self-adjoint operators on 
Hilbert spaces to the Weyl-Segal commutation relation.\\

$\textbf {Theorem 15}$\\
From the Stone-Von Neumann theorem, we can construct the following formula. 
The quantum characteristic function ${\bf G}({\bf z})$ for the class of quantum Gaussian state is given as follows:
\begin{eqnarray}
{\bf G}({\bf z})&=&Tr {\bf U}|\Phi><\Phi| {\bf U}^{\dagger} {\bf V}({\bf z}) \nonumber \\
&=&Tr |\Phi><\Phi|{\bf V}({\cal L}^T {\bf z}) 
\end{eqnarray}
where
\begin{eqnarray}
{\bf V}({\bf z})&=&\exp \{i {\bf R}^T {\bf z}\} \\
{\bf R}&=&[({\bf q}_1,{\bf p}_1), \dots ,({\bf q}_M,{\bf p}_M), ]^T 
\end{eqnarray}
and where $({\bf q}_i,{\bf p}_i)$ are the canonical conjugate operators. Then ${\cal L}$  in Eq(111) is 
called a symplectic matrix and it is given by 
\begin{eqnarray}
&&{\cal  L}=\left (
\begin{array}{ccccc}
r_{11}e^{i\theta_{11}}& \dots &r_{1M}e^{i\theta_{1M}}   \\
r_{21}e^{i\theta_{21}}& \dots  & r_{2M}e^{i\theta_{2M}}\\
\vdots & \dots  & \vdots \\
r_{M1}e^{i\theta_{M1}} & \dots   & r_{MM}e^{i\theta_{MM}}  \\
\end{array}
\right )
\end{eqnarray}
${\bf U}$  in Eq(111) is called  the unitary operator associated with symplectic transformation ${\cal L}$. \\

The protocol is given as follows:\\
\textbf{(i)} A set of ${\cal L}(r_{i,j}, \theta_{i,j})$ with different elements is prepared. \\
\textbf{(ii)} One ${\cal L}$ is selected from the set using a pseudorandom sequence 
with the secret key ${\rm K}_S$. \\
\textbf{(iii)} A quantum ciphertext is generated by a unitary transformation associated with the selected ${\cal L}$ as follows:

Here let us denote a vector of complex amplitudes $\alpha$ of coherent state as follows:
\begin{equation}
\vec{\alpha}_{in}=(\alpha_1, \alpha_2, \dots , \alpha_M)
\end{equation}
then we have the following relation.
\begin{equation}
\vec{\alpha}_{out}={\cal L} \vec{\alpha}_{in}=(\alpha_1^{out}, \alpha_2^{out}, \dots , \alpha_M^{out})
\end{equation}
As a result, the unitary transformation for the coherent state sequence is given  as follows:
\begin{equation}
{\bf U}|\Phi> =|\Phi_{out}> =|\alpha_{1}^{out}>|\alpha_{2}^{out}> \dots |\alpha_{M}^{out}>
\end{equation}
Thus, randomization of complex amplitudes through ${\cal L}$ converts the basic codes of coherent states into 
quantum ciphertext with arbitrary complex amplitude by ${\bf U}$.
When $r_{i,j}=1, \forall i,j$, the above is called phase randomization.

On the other hand, Bob's decryption procedure involves applying the inverse ${\bf U}^{-1}$ of a pseudorandomly 
selected unitary transformation to the quantum ciphertext using the same pseudorandom numbers.
As a result, a receiver with the key can always receive the basic quantum state code signal before the randomization.
For a more detailed explanation, see the literature [56].\\

\subsection{\textbf{Social implementation}}

\subsubsection{\textbf{Development of transceiver for quantum stream cipher}}

Finally, we will explain the development status of transceivers for practical use of the above encryption.
Fig.10 show the history of the development of the tranceiver. 
In 2002, a handmade prototype was developed and communication experiments were successfully conducted at 125Mbps.
 Subsequently, with the cooperation of Panasonic and Hitachi, a two-way prototype transceiver was developed.
Currently, the development of a commercial transceiver for standard quantum stream ciphers has been completed, 
and we are now in the phase of improving it to generalized quantum stream ciphers with several randomizations. 

On the other hand, P.Kumar and his group of Northwestern University developed independently the world's first 
high performance transeiver of quantum stream cipher in 2002, 2003, and 2009 [57, 58, 59]. 
They also demonstrated it for application aircraft communication.
\\

\subsubsection{\textbf{Application to global optical network of 100 Gbit/sec of quantum stream cipher}}

When developing new capabilities in cryptography, it is unacceptable to degrade conventional communication performance. 
To avoid deterioration of communication characteristics, the application of coherent state, 
which is the miraculous quantum state, is essential.
The cryptographic techniques introduced in this section does not degrade communication characteristics.

This encryption device can be operated without changing the structure of the current global optical network system 
with optical amplifier repeaters. In other words, encryption of ultra-high-speed optical communications can be 
completed simply by replacing the current optical transceivers with quantum stream transceivers.

Key distribution for the upper layers of the network utilizes quantum-resistant or PQC, and after key sharing, 
it can be instantly replaced with a new key using quantum stream cipher, 
Thus, we recommend that the global network applications in the real world become schemes shown in Fig.11.
That is, the best solution to defend against today's urgent cyber attacks is to implement combined scheme of 
the quantum-resistant public key cryptography as recommended by NIST and quantum stream cipher to protect data [60,61].
Much research and development is being carried out towards these goals [62,63,64,65,66].

On the other hand, the threat of eavesdropping from undersea cables will become extremely important.
This is because in undersea communication networks that are approximately 10,000 km long, signals are transmitted 
from light to light via optical amplifiers. Furthermore, technology has been established to eavesdrop on 
optical signals from optical amplifier repeaters.

The quantum stream cipher is expected to be a technology that can address this issue 
and preliminary experiments have already been carried out [67]. 
In such systems, keys are pre-installed and no key distribution is required.\\

\subsubsection{\textbf{Business for quantum stream cipher service}}
Commercialization of quantum stream cipher transceivers and their application systems began in the United States in 2025. 
Services that meet customer needs will be available within a few years.

\begin{figure}
\centering{\includegraphics[width=8cm]{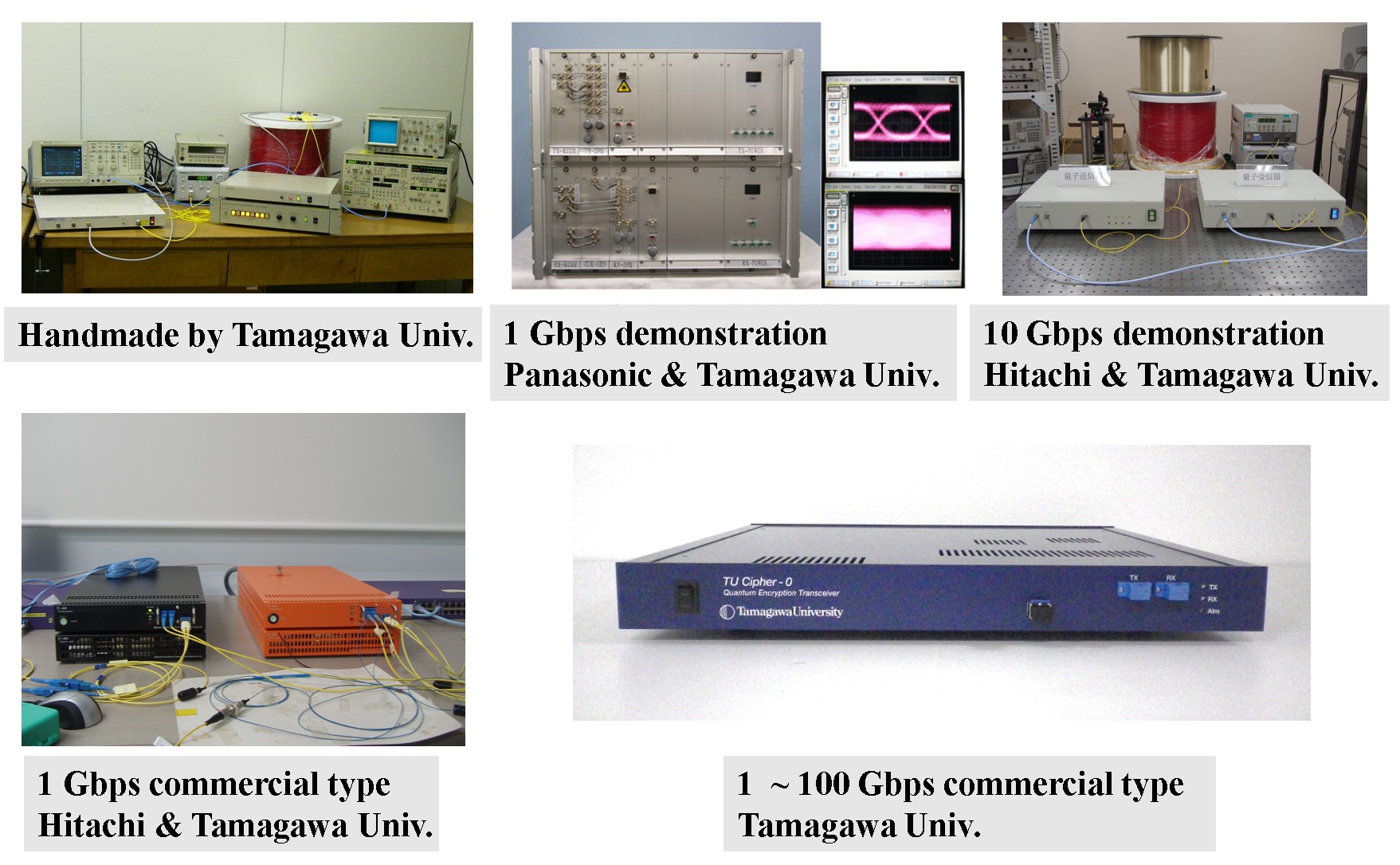}}
\caption{History of the development of quantum stream cipher communication devices: 
Providing encrypted communication of arbitrary ultra-high-speed binary data with randomized coherent state signals without delay.
Setup is complete by simply replacing the optical transceiver currently in use with the above.}
\end{figure}

\begin{figure}
\centering{\includegraphics[width=6cm]{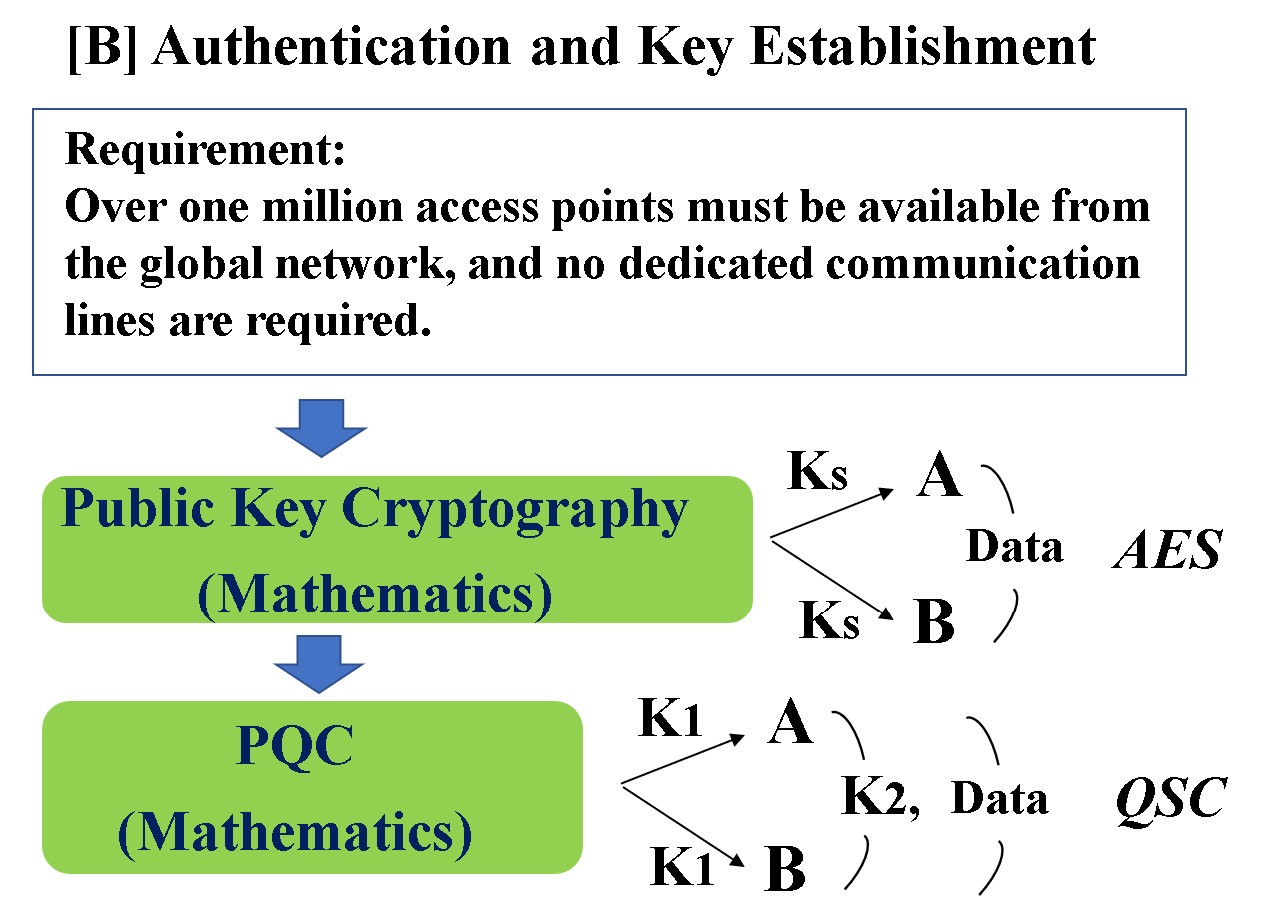}}
\caption{A scheme for total system.
AES is advanced encryption standard and QSC is quantum stream cipher. $K_1$ is instantly replaced by $K_2$ using QSC. }
\end{figure}

 \section{\textbf{Sensor applications beyond the standard quantum limit}}
 
\subsection{\textbf{Bell state based on Entangled Coherent State}}
In general, the Bell states consist of four orthogonal states.
However, the four entangled states based on non-orthogonal states are called 
quasi Bell states.
The specific example is the following states based on coherent states:

\begin{eqnarray}
\left\{
\begin{array}{lcl}
|\Psi_1 \rangle &=& h_{1} (|\alpha \rangle_A|\alpha \rangle_B
+|-\alpha \rangle_A|-\alpha \rangle_B ) \\

|\Psi_2 \rangle &=& h_{2} (|\alpha \rangle_A|\alpha \rangle_B
-|-\alpha \rangle_A|-\alpha \rangle_B )\\

|\Psi_3 \rangle &=& h_{3} (|\alpha \rangle_A|-\alpha \rangle_B
+ |-\alpha \rangle_A|\alpha \rangle_B )\\

|\Psi_4 \rangle &=& h_{4} (|\alpha \rangle_A|-\alpha \rangle_B
-|-\alpha \rangle_A|\alpha \rangle_B)

\end{array}
\right.
\end{eqnarray}
where
$\{h_{i}\}$ are constants for normalization: $h_{1}=h_{3}=1/\sqrt{2(1+\kappa^{2})}$,
$h_{2}=h_{4}=1/\sqrt{2(1-\kappa^{2})}$, and where
$ \langle \alpha | -\alpha \rangle = \kappa$ and
$\langle -\alpha | \alpha \rangle = \kappa^*$.

Some of these quasi Bell states are not orthogonal to each other.
Here, if $\kappa = \kappa^*$, then the Gram matrix of them
becomes very simple as follows:
\begin{equation}
G=
\left(
\begin{array}{cccc}
1& 0& D& 0\\
0& 1& 0& 0\\
D& 0& 1& 0\\
0& 0& 0& 1\\
\end{array}
\right)
\label{ohmsgrammat}
\end{equation}
where $D=\frac{2 \kappa}{1 + {\kappa}^2}$.
Let us employ the entanglement of formation as the degree of entanglement.
 The degrees of entanglement for quasi Bell states becomes as follows:
 \begin{eqnarray}
& &E(|\Psi_1\rangle)=E(|\Psi_3\rangle) \\
                 &=& - \frac{1+C_{13}}{2} \log \frac{1+C_{13}}{2}
                   - \frac{1-C_{13}}{2} \log \frac{1-C_{13}}{2}\nonumber
\end{eqnarray}
where $C_{ij}=|\langle\Psi_i|\Psi_j\rangle|$, 
and we have the special property such as  
$E(|\Psi_2\rangle)=E(|\Psi_4\rangle)=1$ [68].
Thus $|\Psi_2\rangle$ and  $|\Psi_4\rangle$ have the perfect entanglement. 

These physical properties and the mechanisms of teleportation were elucidated by S.J.van Enk [69].\\
 
\subsection{\textbf{Quantum Reading Scheme}}
The term of quantum reading as a sensor was pioneered by Pirandola [70] (Appendix).
We will employ the PSK scheme such that the memory on the classical disk consists of the flat and concave 
which correspond to $``0"$ and $``1"$, respectively.
This type of memory has been invented in the beginning of the 1980's, employing a Gas laser with high coherence 
as the light source. 
Later, its scheme was replaced with a laser diode with high coherence in which the decision scheme consists 
of the effect of reflection wave injection in laser diodes [71].

For the quantum version of such kind of sensor, one can describe the phase shift by an unitary operator as follows:
\begin{equation} 
U(\theta)=exp(-\theta {a}^{\dagger}a)
\end{equation}
where $a$ and ${a}^{\dagger}$ are the annihilation and creation 
operators for bosonic systems, respectively. Here, the phase factor $\theta$ is 
set as between 0 and 2$\pi$.

The reading method in the current PSK scheme is to illuminate 
the laser light (coherent state) on a disk, and to read the phase 
difference. That is, the signal states of light to discriminate are
\begin{eqnarray}
|\alpha(0) \rangle &=& {\it {I}} |\alpha \rangle \\
|\alpha(1)\rangle &=& U(\theta)|\alpha \rangle
\end{eqnarray}
where ${\it {I}}$ is the identity operator.\\

\subsection{\textbf{Error Free Sensor Applicable to Reaction Control}}

Let us introduce an application of the quasi Bell state to reaction control system [72]. 
Here we restrict the problem to the binary detection. So detection targets 
are two quantum states. Since the phase difference will be $\pi$, 
the problem is to read the phase shift $\pi$ from the steady state or 
input state.

The coherent states are prepared in the current classical memory, 
and the target signal model becomes as follows:
\begin{eqnarray}
|\alpha(0) \rangle &=& {\it {I}} |\alpha \rangle=  |\alpha \rangle\\
|\alpha(1)\rangle &=& U(\theta=\pi)|\alpha \rangle = |-\alpha \rangle
\end{eqnarray}
If one employs the conventional homodyne receiver to discriminate 
the above quantum states, the limitation is imposed by so called 
quantum shot noise limit. So this is the standard quantum limit in 
the modern information technology.

However, in general, it is reasonable to employ a quantum optimum receiver
in order to overcome the standard quantum measurement which is achieved 
by a homodyne receiver. In this case, from Eq.(14), the limitation depends 
on the inner product between two quantum states. 
For example, the inner product of two coherent states is
\begin{equation}
\langle \alpha|-\alpha\rangle =exp{(-2|\alpha|^2)}
\end{equation}

Let us assume that  $|\Psi_{2} \rangle$ of quasi Bell states in Eq.(118) 
is employed as the light source, and the $B$ mode is illuminated to 
a memory disk. The reflection effect $U_B(\theta)$ operates on $B$ mode, 
so the channel model is 
\begin{equation}
{\cal {\epsilon}}_{A \otimes B} =I_A \otimes U_B(\pi)
\end{equation}

Then, the target states become as follows:
\begin{eqnarray}
|\Psi_{2}(0) \rangle &=& h_{2} (|\alpha \rangle_A|\alpha \rangle_B
-|-\alpha \rangle_A|-\alpha \rangle_B )  \nonumber \\
|\Psi_{2}(1)\rangle &=& {\cal {\epsilon}}_{A \otimes B} 
h_{2} (|\alpha \rangle_A|\alpha \rangle_B
-|-\alpha \rangle_A|-\alpha \rangle_B ) \nonumber \\
&=&h_{2} (|\alpha \rangle_A|-\alpha \rangle_B
-|-\alpha \rangle_A|\alpha \rangle_B ) 
\end{eqnarray} 
Thus, the input state $|\Psi_2(0)\rangle =|\Psi_2\rangle$ is 
changed to $|\Psi_2(1)\rangle = |\Psi_4\rangle$. 
The inner product between the above two entangled coherent states 
in the quasi Bell state is 
\begin{equation}
\langle\Psi_{2}(0)|\Psi_{ 2}(1)\rangle= 0
\end{equation}
That is, the inner product becomes zero, and it is independent of 
the energy of light source.
This is a  property of the quasi Bell state.
To check this special property, one can examine the different phase 
shift as follows:
\begin{equation}
|\langle\Psi_{2}(0)|I_A \otimes U_B(\theta) |\Psi_{ 2}(0)\rangle |> 0
\end{equation}
where $\theta \ne \pi$.

Finally, let us employ the quantum optimum receiver for the binary 
pure state of two modes. The ultimate detection performances of 
systems with the coherent state and homodyne, coherent state and quantum receiver, 
and quasi Bell state and quantum receiver are given, respectively, as follows:
\begin{eqnarray}
{\bar P}_e(C) &=& {\bar P}e(Homodyne)\\
{\bar P}_e(Q1)&=&\frac{1}{2}[1-\sqrt {1-4\xi_0\xi_1 e^{(-4|\alpha|^2)}}]\\
{\bar P}_e(Q2) &=& 0
\end{eqnarray}
Thus, one can see that the property of the quasi Bell state provides 
an attractive improvement, and this property can be obtained only 
by the combination of the nonclassical state and the  quantum optimum receiver.
Although there are many nonclassical states, almost all are not changed 
from the input state to the orthogonal state to the original input state 
just by reflection. So we interpret that such an effectiveness of 
the quasi Bell state comes from the special phenomena on entanglement 
based on non-orthogonal states, which is a feature of quasi Bell states.\\

\section{\textbf{Conclusion}}

This paper has explained the development of quantum communication theory and its link to Shannon information transmission theory. 
In particular, it shows how quantum gain appears in the performance of functions such as communications designed using 
these theories.
While quantum applications are currently being actively discussed, new mechanisms that sacrifice some of the conventional 
performance when applying quantum properties cannot be applied to real-world situations. 
In other words, quantum communication using qubits, etc., is not suitable for social implementation, 
because such schemes cannot maintain the current communication capabilities.
In addition, we explained that improving delay characteristics (latency) is essential for communication systems in the 21st century.
Therefore, it is essential to advantageously apply quantum effects to modern optical communications, 
which have the highest performance for communication functions. The theory introduced here has been 
shown to be compatible with such system design.
Finally, the appendix discussed paths for future development.
In conclusion, readers of this paper will be able to understand the structure of the design principles of ultra-high speed 
quantum optical communication, quantum stream ciphers and quantum sensors.\\

\section*{\textbf{Appendix}}

\textbf{[A] Towards Social Implementation of Quantum Technology }\\
As telecommunications and microwave communications have evolved into optical communications, a major shift is occurring 
in the way communications theory, which governs their design. In general, there are three fundamental aspects of communications 
such as the amount of information transmitted (bits/sec), error characteristics (reliability), and communication distance. 
Currently, delay characteristic (latency) is also essential. So far, it was considered acceptable 
to sacrifice delay characteristics in order to improve these three fundamental characteristics. 
In particular, for error-correcting codes, a longer code length was not considered a major issue. 
On the other hand, there has been a noticeable trend away from technical theory and towards abstract theory.

To prevent these trend from being carried over into quantum communication theory, Kennedy-Haus proposed training 
researchers with a careful focus on balancing the roles of mathematics, physics, and electronics (Fig.12).
Currently, the quantum Shannon information theory has been developed as a result of this research, 
and communication technologies that apply quantum effects based on this theory are being implemented in society.\\

\begin{figure}
\centering{\includegraphics[width=8cm]{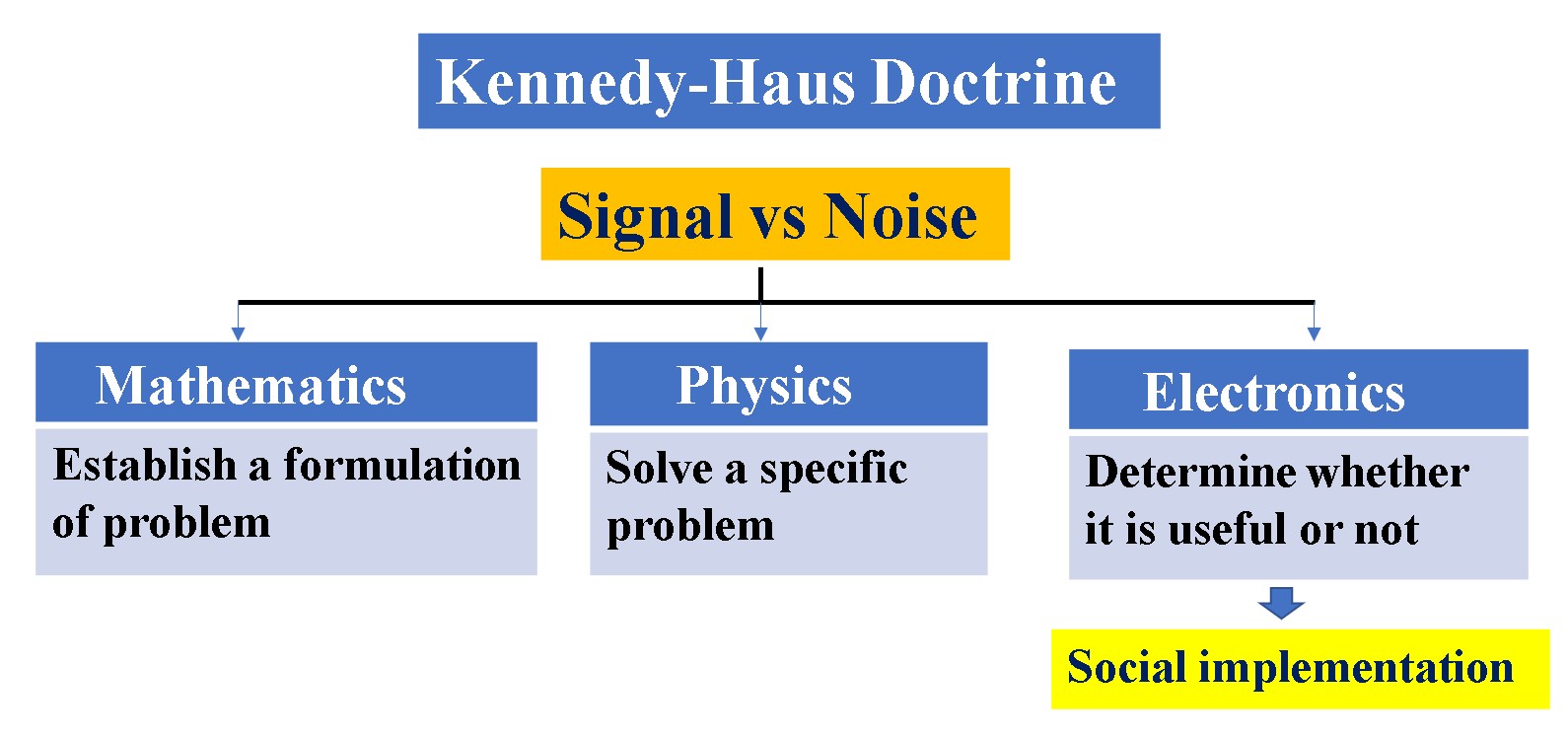}}
\caption{Fundamental concepts for advancing research into utilizing quantum effects for the advancement of communications.}
\end{figure}

\textbf{[B] Towards Further Development of Theory}\\
We will introduce the challenges and notable contributions towards further development of 
quantum communication theory in three fields.\\

\textbf{(i) Mathematics}\\
Entropy, POVM, and its related formulae are simply measures, and their operational meaning is, in general,  
guaranteed separately by various theorems.
Any arbitrary interpretation of the above measures in physics and mathematics is meaningless.
A typical example is Shannon entropy. It first acquires meaning through the source and channel coding theorem.

In the following, we present contributions to further developments regarding the operational meaning 
of measures introduced in quantum theory.\\

{\bf $\cdot$ {\it Holevo-Umegaki}}\\
Von Neumann entropy is given operational meaning as the maximum value of Shannon capacity through the HSW theorem.
On the other hand, The Umegaki relative entropy [73] has operational meaning in the field of quantum Shannon theory 
as a derivative of Holevo theory.

The need for a mathematical generalization of the quantum measurement process was emphasized by Umegaki and others. 
Then P.Benioff, C.W.Helstrom, H.P.Yuen emphasized the operational meaning of the POVM, 
which was proposed as a mathematical generalization.
Finaly the formulation of POVM for quantum communication theory was systematized by Holevo, 
following the concept of decision operator interpretation by Helstrom.
Their operational meaning forms the basis of quantum communication theory.\\

\textbf{$\cdot$ {\it Davies-Ozawa}}\\
E.Davies has published a book that provides an overview of the mathematical background of the quantum measurement process [74].
Then he defined a concept of Instrument as follows:\\

{\bf Definition 12 }\\
Let $\Omega$ be a set of with a $\sigma$-field ${\cal F}$. Then one can define the POVM on $\Omega$ to be a map 
${\cal E}$: ${\cal F} \rightarrow {\cal L}^+(V)$.

If POVM satisfies the further condition
\begin{equation}
Tr[{\cal E}(\Omega)\rho]=Tr[\rho], \quad \forall \rho \in V
\end{equation}
then it is called ``instrument".\\

 M.Ozawa [75, 76, 77] took the debate between Yuen and Caves in the theory of gravitational wave detection 
 as an opportunity to develop Davies's instrument theory to a high level, including ``Ozawa's representation theorem". 
 It is expected that his results will help to clarify the operational meaning through the theory of gravitational wave 
 detection in the future.\\

\textbf{$\cdot$ {\it Belavkin}}\\
V.P.Belavkin formulated the mathematical foundation for extending the classical linear and nonlinear filtering theory 
to quantum systems [78,79].
For example, he has developed quantum versions of Kalman-Bucy type linear filters and Bellman equations 
for control systems with feedback.
The mathematical basis for this is ``quantum stochastic calculus in Fock space and stochastic equation".
\begin{eqnarray}
|d\psi_t >&=&\frac{1}{i\hbar}H(u_t)|\psi_t >dt \nonumber \\
&-&\frac{1}{2}\sum_{\alpha}[L^{\dagger}_{\alpha}L_{\alpha}
-2\lambda_{\alpha}(t)L_{\alpha}+\lambda^2_{\alpha}(t)]|\psi_t >dt \nonumber \\
&+&\sum_{\alpha}[L_{\alpha}-\lambda_{\alpha}(t) ]|\psi_t > dW_{\alpha}(t)
\end{eqnarray}

The main results may provide theory of optimal estimation of quantum measurements, 
which is essential for future development of quantum control theory.

In sum, new mathematical research with operational meaning will provide the impetus for further development 
of quantum Shannon theory.\\

\textbf{(ii) Physics}\\
Glauber's coherent state theory opened up the field of quantum optics. 
Yuen pioneered the theory of two-photon coherent states to control uncertainty relation between quadrature amplitude [80]. 
This opened up new technological possibilities for the controllability of quantum states of light. 
Currently, these are being developed in various fields as squeezed states.

Meanwhile, the theory of entanglement of quantum states mentioned above was proposed by Sanders [81].
Its operational meaning was explored through theories by van Enk [69] and others, and applications were opened up.
Furthermore, P.Grangier [82] and H.Jeong [83,84] made significant contributions to its analysis of physical properties 
and experimental development.\\

\textbf{(iii) Detection and estimation theory}\\
In this section, we introduce several extension of the quantum detection and estimation for communication theory.\\

\textbf{$\cdot$ {\it Eldar}}\\
In the field of electrical signal processing, squared error is sometimes used to evaluate the error. 
She introduced squared error evaluation for finite alphabet as an system evaluation for quantum signal processing [85].
The evaluation measure is given as follows:
\begin{equation}
\Delta =\sum_{i=1}^M <e_i|e_i>, \quad |e_i >=|\psi_i > - |\mu_i>
\end{equation}

Eldar has formulated an optimum theory by considering quantum states as signal vectors.
Thus, she has assumed pure-state ensembles $\{|\psi_i>\}$ and seeked a POVM $\Pi_i =|\mu_i ><\mu_i |$ consisting of 
rank-one positive operators with measurement vectors that minimize the sum of the squared norms of the error vectors. 
However, the challenge is to clarify the meaning of POVM, which is the solution.\\

\textbf{$\cdot$ {\it Chefles-Barnnet}}\\
Chefles-Barnnet formulated an optimum theory of `` Unambiguous decision" [86]. 
This combines error free discrimination with a certain fraction of inconclusive results. 
They showed that a necessary and sufficient condition for the existence of unambiguous measurements for distinguishing
 between $M$ quantum states.\\

\textbf{$\cdot${\it  Pirandola}}\\
He proposed a new detection concept in quantum detection theory that is called ``quantum reading" [70].
The quantum reading is defined such that 
it is the use of input quantum resources (such as entanglement) and output quantum measurements to enhance the retrieval 
of classical information stored in the cells of a memory.

His idea developed to  quantum channel discrimination (QCD) [87]. 
The QCD is a very general problem where an ensemble of quantum channels is prepared in a black box, and descriminate them.

The several generalizations of the unambiguous decision and channel discrimination has been developed by K.Nakahira [88,89,90].
\\

\textbf{$\cdot$ {\it Personick}}\\
He is a pioneer in research into the specific structure of modulation and demodulation based on quantum estimation [91].
Fundamentally, estimation theory deals with analog signals.
Therefore, the evaluation function is the minimum mean square error as follows:.
\begin{equation}
\min_{{\bf X}}{\cal C}({\bf X})=\int \xi(s)[Tr \rho(s)({\bf X}(ds) - sI)]^2ds
\end{equation}
He found the optimum condition for the above problem and his work led to the filter theory.

R.O.Harger and his group attempted to develop a quantum linear filter using Personick's theory [92].
These operational meanings are clear and provide a basis for further development of estimation theory.\\

\textbf{[C] Quantum Shannon Information Transmission Theory}\\
Since the publication of Shannon's theory, research into the construction of fundamental theories and 
their applications has expanded into an extremely broad field. 
The true essence of this theory lies in coding theory. 
However, research in the second half of the 20th century focused on abstract theories and mathematical research 
into unrealizable asymptotics, which Shannon himself was not favorable.
Coding techniques are used in actual communications technology,
but delay (latency) issues as their greatest drawback have become a serious problem in the 21 century. 

Rather than theories that achieve communication capacity, 
there is growing demand for theories that can keep code lengths below a certain level when the rate is fixed 
below the cut-off rate. 

Meanwhile, this type of research into quantum systems remains limited to asymptotic discussions 
that stick to old styles. 
This is  inherently useless. 21st century communications are moving towards the realization of 
ultra-high-speed, low-latency communication systems. 
The research of quantum information community must meet these demands.

Considering such a situation, the development of reliability function theory is appropriate.
Dalai's work [93] deserves great recognition. He provided the fundamentals of 
random coding, spher-packing, and expurgation. Kato [94] analyzed specific examples of coherent state signals 
in real-world ultra-high-speed communications and clearly demonstrated their properties based on Dalai's theory. 
Code constructions that take into account the properties of specific quantum states have already been considered [95].

We recommend the following books to help you learn these basic theories such interdisciplinary fields
 [96,97,98,99,100,101, 102, 103]. \\

1. R.L.Stratonovich, ``Theory of information and its value".

2. R.G.Gallager, ``Principle of digital communication".

3. G.C.Papen and R.Blahut, ``Lightwave communications".

4. J.Perina, ``Quantum statistics of linear and nonlinear optical phenomena".

5. I.B.Djordjeviv, ``Quantum communication, quantum networks and quantum sensing".

6. G.Cariolaro, ``Quantum communications".

7. L.Cohen, H.V.Poor, and M.O.Scully (editor), ``Classical, semiclassical and quantum noise".

8. M.Ohya and D.Petz,`` Quantum entropy and its use".
\\

\textbf{[D] Towards the Development of Quantum Cipher for Global Optical Network}\\
Generally, the characteristics of cryptography are evaluated based on many factors, including security, 
scalability, implementation characteristics, latency characteristics, cost, and compatibility with a real systems. 
Unless a balance between these factors is maintained, social implementation is impossible.

When evaluating the characteristics of physical cryptography, communication characteristics are particularly important. 
Therefore, using single photons or quantum entanglement is impractical.

Future physical cryptography that utilizes quantum properties must prioritize protecting the data used in today's 
ultra-high-speed optical communications. To achieve this, quantum stream cipher, which can encrypt data 
while maintaining ultra-high speeds, would be the best solution. This is because it follows the trend of 
modern cryptography represented by Massey  and others [104, 105, 106,107] and provides new security features.

\end{document}